\documentclass[12pt]{article}

\pdfoutput=1

\usepackage{amsmath}
\usepackage{amssymb}
\usepackage{graphicx} 
\usepackage{natbib}

\usepackage[margin=1in]{geometry}

\usepackage[utf8]{inputenc}
\usepackage{placeins} 
\usepackage{setspace} 

\usepackage[font=small,labelfont=bf, width=0.90\textwidth]{caption}

\usepackage{multicol}
\usepackage{pdfpages}

\title{Not even wrong: The spurious link between biodiversity and ecosystem functioning}
\author{ \normalsize{$^{a1}$Pradeep Pillai, $^{1}$Tarik C. Gouhier}\\
\normalsize{$^{1}$Marine Science Center, Northeastern University,}\\
\normalsize{430 Nahant Rd, Nahant, MA 01908,}\\
\normalsize{$^{a}$To whom correspondence should be addressed. E-mail: pradeep.research@gmail.com} }

\date{}

\begin{document}

\linespread{1.0} 


{\centering

\textbf{\Large{Not even wrong:} }\\ \textbf{\large{The spurious link between biodiversity and ecosystem functioning}} \\
\bigskip
Pradeep Pillai\textsuperscript{a1} , Tarik C. Gouhier\textsuperscript{1} \\
\textit{
\textsuperscript{1}Marine Science Center, Northeastern University,\\
30 Nahant Rd, Nahant, MA 01908\\
\textsuperscript{a} To whom correspondence should be addressed. E-mail: pradeep.research@gmail.com \\
}
}


\begin{abstract}
  \noindent Resolving the relationship between biodiversity and ecosystem functioning has been one of the central goals of modern ecology. Early debates about the relationship were finally resolved with the advent of a statistical partitioning scheme that decomposed the biodiversity effect into a ``selection'' effect and a ``complementarity'' effect. We prove that both the biodiversity effect and its statistical decomposition into selection and complementarity are fundamentally flawed because these methods use a na\"ive null expectation based on neutrality, likely leading to an overestimate of the net biodiversity effect, and they fail to account for the nonlinear abundance-ecosystem functioning relationships observed in nature. Furthermore, under such nonlinearity no statistical scheme can be devised to partition the biodiversity effects. We also present an alternative metric providing a more reasonable estimate of biodiversity effect. Our results suggest that all studies conducted since the early 1990s likely overestimated the positive effects of biodiversity on ecosystem functioning.
\end{abstract}

\bigskip
\def\undertilde#1{\mathord{\vtop{\ialign{##\crcr
$\hfil\displaystyle{#1}\hfil$\crcr\noalign{\kern1.5pt\nointerlineskip}
$\hfil\widetilde{}\hfil$\crcr\noalign{\kern1.5pt}}}}}

 \section*{Introduction}
The notion that increasing biodiversity will enhance the value of some aggregate ecosystem property (i.e., ``ecosystem functioning'') has now achieved the status of a near truism in ecology. Despite early debates over its legitimacy \citep{huston_97,jocelynkaiser_00,huston_00} the principle of a positive biodiversity-ecosystem functioning (BEF) relationship has been consistently affirmed in ecological studies for the last quarter-century \citep{tilman.knops.ea_97,naeem.thompson.ea_94, hector_99, hooper.chapin.ea_05}, while at the same time being subject to surprisingly little conceptual or theoretical challenge to its underlying premises and methods. Unfortunately, we demonstrate that all current BEF frameworks rest on several critical flaws ranging from a trivial quasi-circularity inherent in the approach, to the unexamined effects of nonlinearity which, when combined, inflate the positive effect of biodiversity on ecosystem properties.

The positive BEF relationship often observed in studies is usually obtained by measuring ecosystem functioning in communities relative to null expectations based on neutral or zero-sum game assumptions, whereby all species have the same fitness and thus equally share a common niche. Although neutral theory has shown that such fitness equalizing mechanisms can promote coexistence \citep{hubbell_01,chesson_00a} it is now widely recognized that niche partitioning is common in nature and that coexistence is often driven by fitness stabilizing mechanisms that increase intraspecific competition relative to interspecific competition \citep{chesson_00a,adler.ellner.ea_10,levine.hillerislambers_09}. Indeed, one of ecology’s most fundamental rules, the competitive exclusion principle, posits that long-term coexistence requires some form of niche partitioning \citep{chesson_00a,gause_34}. Although neutral theory can serve as a good null model for studying community structure \citep{rosindell.hubbell.ea_11}, we argue that the assumption of neutrality is not appropriate in BEF studies because it is too strong or na\"ive, and tantamount to a strawman argument. Indeed, by adopting neutrality as a null expectation, BEF studies of non-neutral or niche-structured communities will tend to artificially inflate the positive effect of biodiversity on ecosystem functioning.

The near-axiomatic role of the competitive exclusion principle in ecology means that, knowing nothing else, our default expectation (\textit{ceteris paribus}) would be for species in mixtures to coexist by some form of niche partitioning, and that consequently, any aggregate property (e.g., biomass) that is positively associated with species abundance should naturally increase relative to a naive null expectation that assumes species equally and neutrally share the niche -- that is, some degree of `overyielding' of ecosystem properties should be a natural outcome of coexistence. Thus, there is a strongly circular and trivial element associated with the implicit definition of a measure like ecosystem functioning: much of the functioning measured in BEF experiments likely represents, at least in part, a sort of redundant measure of coexistence amongst coexisting species. The presence of a positive relationship in most BEF studies is thus unsurprising and largely trivial. An important question we should be asking is to what degree are we simply measuring the \textit{coexistence of coexisting species}; that is, to what degree are current BEF measures simply a trivially redundant measure of coexistence? Below we offer a more realistic and useful null expectation based on niche partitioning that can serve as a starting point in the development of more meaningful implicit measures of ``ecosystem functioning''.

In addition to this logical circularity in the underlying premise, there has also been a fundamental mathematical flaw in the theoretical foundation upon which most of the BEF research program has been built over the last two decades: the Loreau-Hector (LH) statistical partitioning scheme \citep{loreau.hector_01}. This scheme \citep{loreau_98a} partitions the net biodiversity effect (the change in aggregate ecosystem properties observed in mixtures relative to null expectation based on the average of all species' monoculture yields) into what Loreau and Hector refer to as a ``selection effect'' and ``complementarity effect''.

The selection effect is the covariance between all species monocultures, $M$, and the change in the proportion of this monoculture that is observed in mixtures relative to expected, $\Delta p$: $n\mathrm{Cov}[M, \Delta p]$, for an $n$-species mixture. This selection effect purportedly measures the degree to which ``species with higher-than-average monoculture yields \textit{dominate the mixtures}'' \citep[][emphasis added]{loreau.hector_01}. The complementarity effect on the other hand is the product of the average monoculture yields and the average proportional changes, $n\overline{M}\;\overline{\Delta p }$, and purportedly measures, as Loreau and Hector claim, ``any change in the average relative yield in the mixture, whether positive (\textit{resulting from resource partitioning or facilitation}) or negative (\textit{resulting from physical or chemical interference})'' \citep[][emphasis added]{loreau.hector_01}.

Over the last two decades, the LH partitioning scheme has been used extensively to quantify and partition the biodiversity effect in experimental studies \citep{cardinale.srivastava.ea_06,cardinale.duffy.ea_12a}. The method has also been extended to further partition the ``selection effect'' \citep{fox_05}, and understand BEF in food web networks \citep{barnes.jochum.ea_18}, as well as across spatial scales \citep{isbell.cowles.ea_18}. However, despite its popularity and many extensions, we show that the claim that the LH partitioning scheme is capable of discerning and measuring the relative roles of selection and complementarity is in general incorrect. In fact, the LH partitioning scheme only holds in the special (and unlikely) case that all species' ecosystem-abundance relationships are perfectly linear in monocultures.

The idea behind this flaw can be grasped intuitively by considering a simple univariate analogy, where an observed property $y(x)$ (e.g., ecosystem functioning measured as biomass) is a function of a single underlying variable $x$ (e.g., biodiversity). For any given change in the observed property, $\Delta y$, it is clear that only if $y(x)$ is a linear function can we claim that
\begin{flalign}
  \Delta y &= \Delta x \, \frac{\mathrm{d}y}{\mathrm{d}x}. &{} \notag
\end{flalign}
Now if we wished to `partition' the total property change $\Delta y$ into two proportions $p_1$ and $p_2$ that add up to 1, then linearity will allow us to state unequivocally that $\Delta y = (p_1\,\Delta x) \,{\mathrm{d}y}/{\mathrm{d}x} + (p_2\,\Delta x) \, {\mathrm{d}y}/{\mathrm{d}x}$, or alternatively, that $p_i \,\Delta y = (p_i\,\Delta x) \,{\mathrm{d}y}/{\mathrm{d}x}$, for any proportion $p_i$. This latter expression clearly indicates that any shifts measured at the observed property level, $\Delta y$, can now be attributed \textit{solely} to corresponding shifts in the underlying variable $\Delta x$. In other words, phenomenological observations can be meaningfully used to make inferential statements about explanatory causes, but only under conditions of linearity.

Although our arguments regarding the LH partitioning can be considered in some ways as a more involved, multivariate version of this illustrative analogy, the intuitive understanding of the underlying flaws from the simple univariate example above holds in the more complicated BEF case examined rigorously below. What is important to note is that the flaw we explore in the LH partitioning method is foundational. It carries over to every partitioning scheme that is ultimately based on it. Even worse, all extensions of the LH scheme published over the last decade or so that simply involved breaking down existing effects into smaller partitions \citep{fox_05,isbell.cowles.ea_18} are likely to have only amplified existing errors.

The (very simple) vector or multivariate calculus framework we develop for exploring the logical and mathematical flaws at the heart of the BEF research program also conveniently provides a geometric interpretation of ecosystem change that can offer an easy tool for visualizing the changes tracked by mathematical expressions, and thus may provide a more intuitive understanding of the mathematical arguments being made. Although this vector and visual approach will likely be new and unfamiliar to some ecologists, we believe its value will quickly become apparent, not only as a useful framework for looking afresh at the approach and premises of the BEF research program, but also for quickly highlighting other new and previously unforeseen flaws that have eluded BEF researchers for decades.

\section*{Theoretical Framework}
\begin{figure}[h!]
  \centering
\includegraphics[scale=0.33]{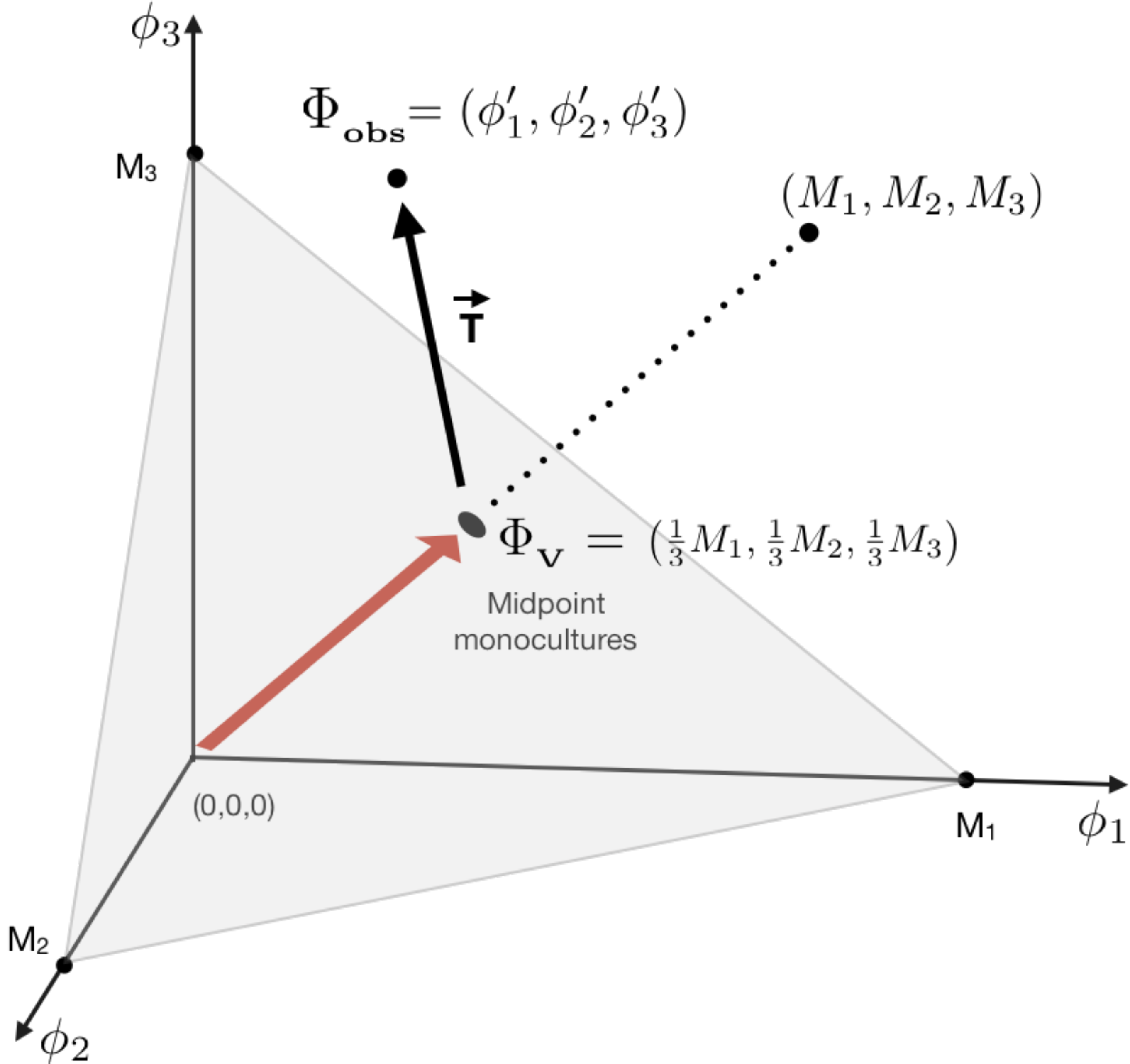}
\caption{Changes in ecosystem properties as changes in state space. Each axis defining the space represents the ecosystem contribution of a given species, $\phi_i$. Example shown for a three-species system. In BEF studies the centroid (center of mass) of the simplex (the surface connecting all the monoculture yields) represents the expected state of the ecosystem, $\Phi_{_\mathbf{V}}$, arising from variation in species growth rates. The difference in the total ecosystem property between the expected and the observed state along the transformational vector $\mathbf{T}$ is the `net biodiversity effect' measured in standard BEF experiments, $\Delta \phi_{_\mathbf{T}}$.}
\label{Fig:fig1}
\end{figure}

Our approach is to visualize ecosystem changes as movement through the state space defined by the ecosystem contribution of each species, where the current state of the ecosystem will be given by the coordinates in the space defined by axes measuring each species' ecosystem contribution: $\phi_1, \phi_2, \dotsc$, etc. If $\boldsymbol{\Phi} = (\phi_1, \dotsc, \phi_n)$ is a positional vector (a vector from the origin to a given point) representing the coordinates in space giving the ecosystem value of each species, then the total ecosystem value at that point is simply the scalar-valued function $\phi(\mathbf{\Phi})=\sum_i^n \phi_i =\phi_1+\phi_2+\dotsb+\phi_n$ (Fig. \ref{Fig:fig1}).

Any total ecosystem change can be partitioned into the effects arising from changes due to \textit{variational} and \textit{transformational} change in the system \citep[\textit{sensu amplio},][]{lewontin_85}. If the ecosystem value at $\boldsymbol{\Phi_\mathbf{V}}$ represents the \textit{expected} value after a single time step due to differential growth of each species, $\phi_{_\mathbf{V}}$ (the \textit{variational} component of change), then the difference in the ecosystem property along the displacement vector $\mathbf{T}$ (Fig. \ref{Fig:fig1}) between this expected value and that of the observed state $\phi_{\mathbf{obs}}$, represents an additional shift due to the \textit{transformational} changes (i.e., $\Delta\phi_{_\mathbf{T}} = \phi_{\mathbf{obs}} - \phi_{_\mathbf{V}}$) that are not simply reducible to variation in species growth (e.g., ecological interactions, environmental effects). In an $n$-species community, the final observed ecosystem value $\phi_{\mathbf{obs}}$ can be expressed as
\begin{flalign} \label{eq:Price1}
\phi_{\mathbf{obs}} &= \phi_{_\mathbf{V}} + \Delta\phi_{_\mathbf{T}} &{}
\end{flalign}

This is simply a modified form of the Price equation \citep{price_70} with the monoculture yields serving as proxies for fitness or expected growth (see Supporting Information 1). The ecosystem functioning quantified in standard BEF experiments is a measure of ecosystem change along this transformational component, $\Delta\phi_{_\mathbf{T}}$, and is referred to as the \textit{net biodiversity effect}. Under the assumptions of BEF experiments, where initial species densities are all equal, we expect the reference point from which we measure ecosystem functioning to be $\mathbf{\Phi_{_V}}=\frac{1}{n}(M_1,\dotsc,M_n)$, which is located at the centroid of the simplex or hyperplane connecting all monoculture yields (Fig. 1). The total ecosystem value at this point is the average of all monocultures, $\phi_{_\mathbf{V}}=\overline{M}$.

This framework for tracking communities and ecosystems through state space can provide a powerful visual tool for relating the statistical and mathematical expressions used in BEF research to the community and ecosystem changes observed in experiments. Below we will use it to investigate the logical and mathematical nature of scientific inference within the BEF research program. In doing so, the framework will help us to elucidate both the problematic logic of the BEF approach to measuring ecosystem functioning (due to its built-in circularity), and the fatal mathematical flaws underlying the Loreau-Hector partitioning scheme and its measurement of the net biodiversity effect.

\section*{Results and Discussion}
\subsection*{Biodiversity effects as redundant measures of coexistence}
A standard operating principle in community ecology is that coexistence is expected when the effects of intraspecific interactions outweigh those of interspecific interactions \citep{chesson_00a,adler.ellner.ea_10,gause_34} leading to species abundances in mixtures likely appearing above the simplex line or plane connecting all carrying capacities. If a property such as biomass is assumed to be directly related to species abundance, then we would also expect that the aggregate property for coexisting species in a mixture will appear above the simplex plane connecting all monoculture yields, and on average, for communities with comparable monoculture yields to exhibit a positive biodiversity effect on ecosystem functioning, $\Delta\phi_{_\mathbf{T}}>0$. Hence, an element of circularity is inherent in all BEF studies because measuring an increase in an ecosystem property that can reasonably serve as a proxy for abundance (e.g., biomass) is simply a roundabout way of measuring the very conditions necessary for coexistence. Coexistence alone will often be sufficient to produce a positive biodiversity effect on ecosystem functioning.

In order to avoid such trivial biodiversity effects, we need to at least account for that portion of the increased ecosystem functioning observed in mixtures that is merely the consequence of species coexistence. A more reasonable starting point for BEF studies would be to use pairwise interactions because the phenomenological effects that species have on each other in pairwise mixtures provide a better starting point to measure the potential effects of diversity by allowing us to discount the portion of the change in an ecosystem property that is merely a redundant measure of coexistence. For a given set of species, the pairwise interaction coefficients, $\alpha_{i,j}$ and $\alpha_{j,i}$, for any two interacting species $i$ and $j$ can be found by solving
\begin{flalign}
  \begin{pmatrix}
  1 & \alpha_{_{i,j}} \\
  \alpha_{_{j,i}} & 1
 \end{pmatrix}
 \begin{pmatrix}
 \phi_{i} \\
 \phi_{j}
 \end{pmatrix}
 &= \begin{pmatrix}
  M_{i} \\
  M_{j}
\end{pmatrix} &{}
\end{flalign}

\begin{figure}[h!]
  \centering
\includegraphics[scale=0.52]{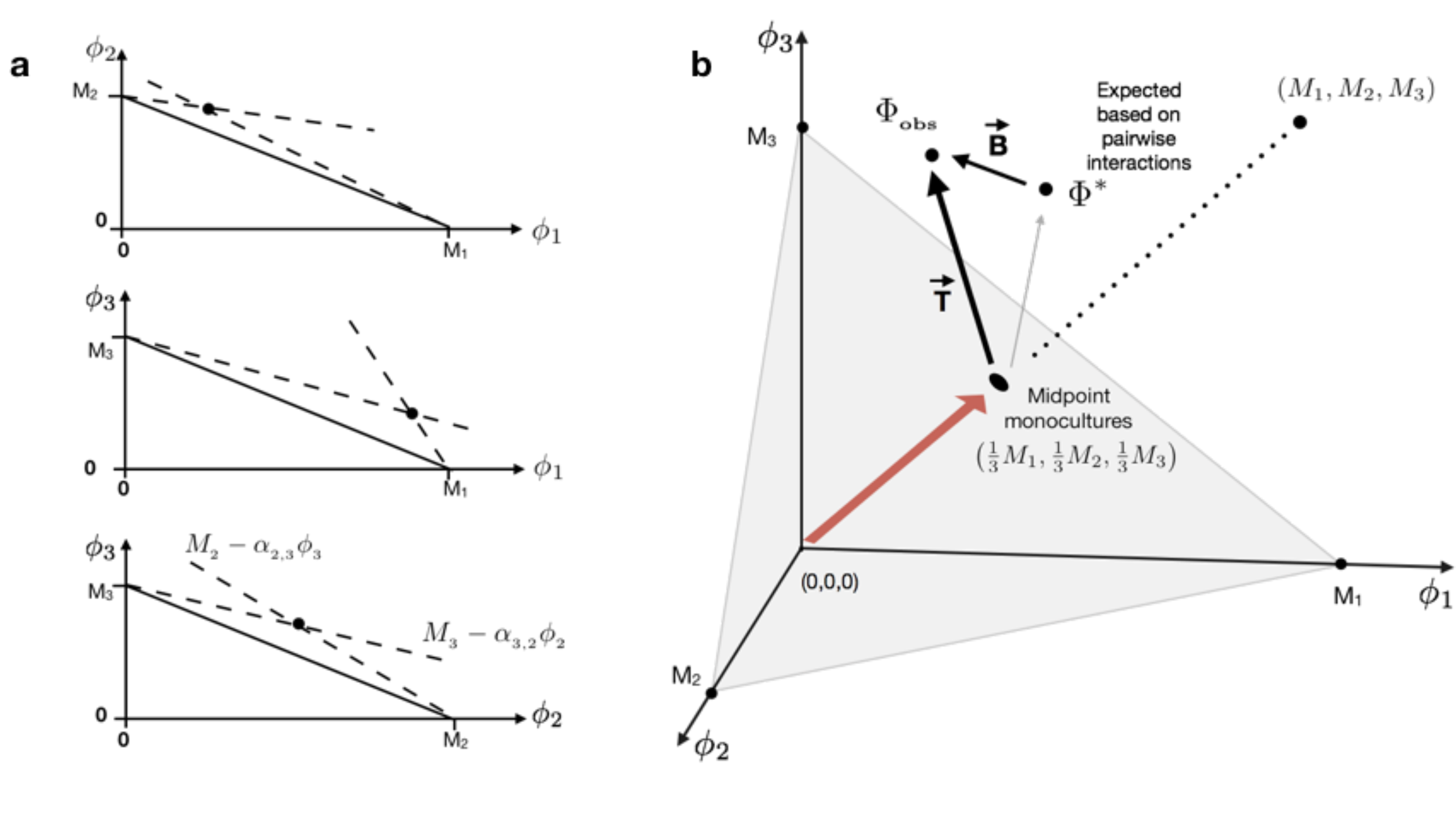}
\caption{Measuring biodiversity effects using pairwise interactions. (a) Conceptual example of 3-species experiment giving pairwise interaction effects. (b) Linearly adding pair-wise interactions give the expected ecosystem state when no higher order interactions operate, $\Phi^*$. Difference in ecosystem property along displacement vector $\mathbf{B}$ between observed, $\Phi_{_\mathbf{obs}}$, and $\Phi^*$ states gives biodiversity effect $\Delta \phi_{_\mathbf{B}}$.}
\label{Fig:fig2}
\end{figure}

If the ecosystem contribution of any species $i$ in more diverse communities results from summing up the phenomenological effects that all other species have had on $i$ in pairwise mixtures, then the expected ecosystem contribution of $i$, $\phi_i^*$, is simply its monoculture yield minus the linear sum of the phenomenological effects of all other species in the community: $\phi_i^* =M_i -\sum_{j\neq i}^n \alpha_{i,j}\phi_j$. By constructing an $n\times n$ community matrix from the pairwise interaction coefficients \citep{macarthur.levins_67a}, $A_{n\times n}=\left\{\alpha_{i,j}\right\}$ we can predict the ecosystem state of higher diversity $n$-species communities ($\mathbf{\Phi^*}=(\phi_1^*,\phi_2^*, \dotsc, \phi_n^* )$) by solving
\begin{flalign}\label{eq:prediction}
  \undertilde{\Phi}^*  &= A^{-1}\undertilde{ M}, &{}
\end{flalign}
where $ \undertilde{\Phi}^* =(\phi_1^*,\phi_2^*,\dotsc, \phi_n^*)^T$ and $ \undertilde{M} =(M_1,M_2,\dotsc, M_n)^T$. Equation \eqref{eq:prediction} gives a baseline prediction based on the minimal assumption that each species' ecosystem effect is simply the linear scaling up of the effects observed in pairwise mixtures (Fig. \ref{Fig:fig2}). Observed departures from this baseline expectation, ($\mathbf{B} = \mathbf{\Phi_{_{obs}}}-\mathbf{\Phi^*}$; Fig. \ref{Fig:fig2}b), indicate that the aggregate ecosystem property of a community is likely determined by higher order interactions between species, or other previously unaccounted for nonlinear effects.

The ecosystem change along the vector $\mathbf{B}$, given by $\Delta\phi_{_\mathbf{B}}$, is a measure of the biodiversity effect beyond that expected due to coexistence, and provides a more meaningful measure of the effects of increasing biodiversity on ecosystem properties than the net biodiversity effect $\Delta\phi_{_\mathbf{T}}$ (i.e., the transformational component of the Price equation). This is because we expect species coexistence by its very nature to entail some form of niche partitioning, and thus, that the final ecosystem state will appear above the simplex, and on average, when monoculture yields are comparable, for $\Delta\phi_{_\mathbf{T}} >0$. Thus, using $\Delta\phi_{_\mathbf{T}}$ is likely to artificially inflate any perceived positive ecosystem effect. In contrast, using $\Delta\phi_{_\mathbf{B}}$ allows us to account for these default expectations by measuring ecosystem shifts relative to a baseline null expectation that accounts for the positive effects that coexistence in itself is likely to have on aggregate ecosystem properties. In other words, given that some overyielding of ecosystem properties is a likely consequence of coexistence by virtue of the competitive exclusion principle (all things being equal), then conducting experiments where one simply assembles community mixtures from species that are known to coexist in nature should require measures that account for the default biodiversity effects already associated with coexistence itself.

To determine the degree to which the naïve assumption of neutrality inherent in BEF studies inflates the biodiversity effect, we performed a series of numerical simulations using the average and the variances of the pairwise interaction coefficients obtained from the BIODEPTH BEF experiment \citep{spehn.hector.ea_05} to generate a series of randomly assembled communities (1000 randomly assembled communities sampled from the species pool of 1000 species) under each biodiversity level (diversity levels: 2, 3, 4, 8, 11, 12, 14) and measured the corresponding ecosystem functioning using both the neutral baseline expectation $\Delta\phi_{_\mathbf{T}}$ and the more realistic expectation based on pairwise mixtures $\Delta\phi_{_\mathbf{B}}$ (see Appendix S2 for details). The simulations provide a proof-of-concept allowing us to more easily demonstrate how a pairwise approach for measuring biodiversity effects can be implemented, as well as to demonstrate how the measured averages and variances of pairwise interaction coefficients from BIODEPTH likely imply that the net biodiversity effects measured in this system involve a significant degree of redundancy or triviality.

\begin{figure}
  \centering
    \includegraphics[width=0.55\textwidth]{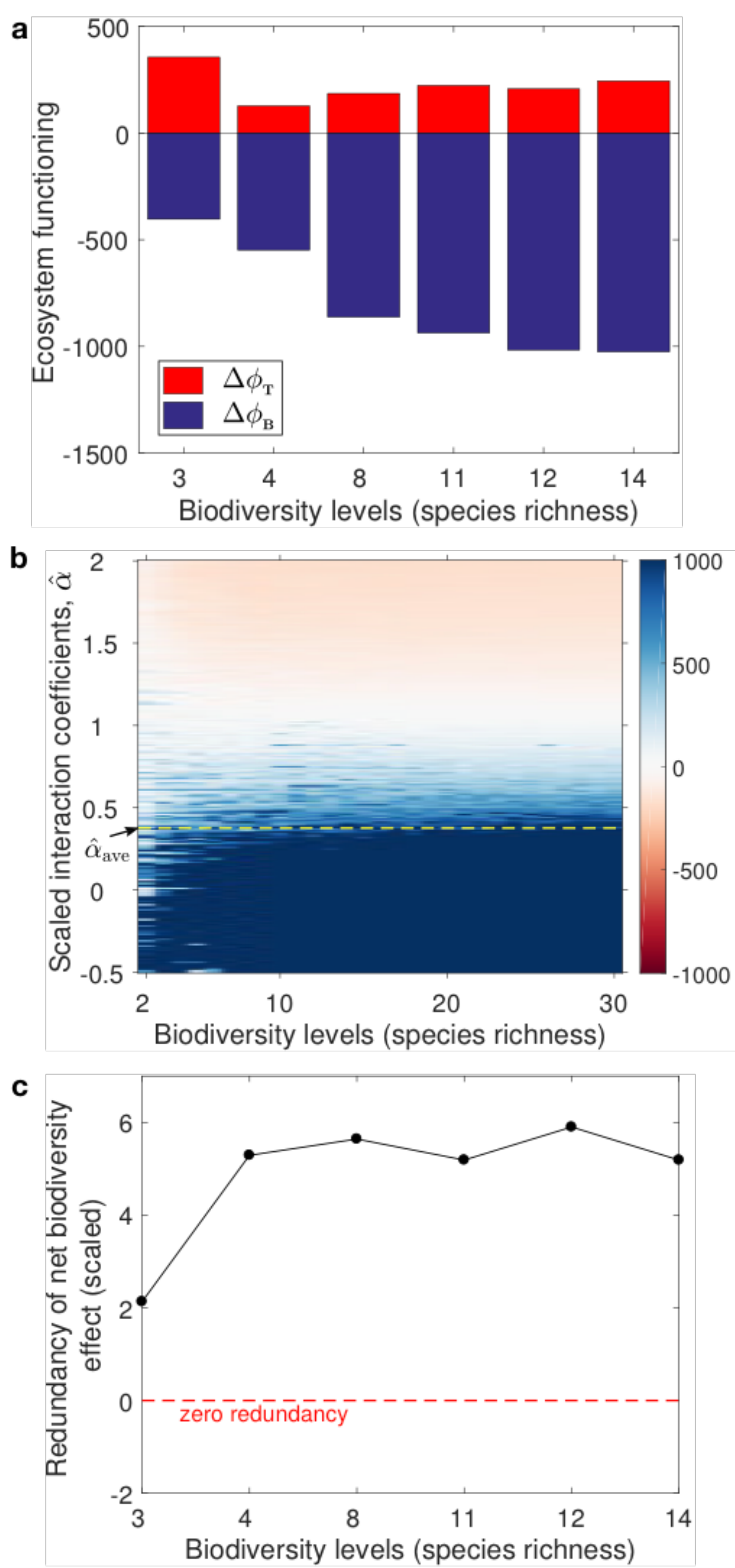}
    \caption{Comparing biodiversity effects in BIODEPTH experiments. (a) Simulations using BIODEPTH data show how $\Delta \phi_{_\mathbf{B}}$ may compare to the net biodiversity effect measured in BIODEPTH, $\Delta \phi_{_\mathbf{T}}$.  (b) Heat map of simulated net biodiversity effects, $\Delta \phi_{_\mathbf{T}}$, for simulations run across a range of $\hat{\alpha}$s and diversity levels (Mean and standard deviation of $\hat{\alpha}$ from BIODEPTH are $\mu = 0.3528$ and $\sigma = 0.6822$; See Appendix S2). (c) Estimates, using simulation data, of the redundancy in measured net biodiversity effects from BIODEPTH experiments. Redundancy is zero (dashed horizontal line) when co-existence accounts for none of net biodiversity effect.}
    \label{Fig:fig3}
\end{figure}

The simulations showed that across all levels of diversity, the effect of biodiversity on ecosystem functioning based on the neutral baseline expectation  $\Delta\phi_{_\mathbf{T}}$ (net biodiversity effect) is systematically positive, whereas the effect of biodiversity on ecosystem functioning based on scaling-up from pairwise mixtures $\Delta\phi_{_\mathbf{B}}$ is always negative (Fig. \ref{Fig:fig3}a). Hence, although the traditional metric based on neutrality would suggest that diversity systematically promotes ecosystem functioning, our metric based on departures from the pairwise mixtures suggests that higher order species interactions, or other nonlinear effects, associated with diversity negatively impact community-level properties and systematically yield negative biodiversity effects that erode ecosystem functioning.

Since the measured average and variance of the BIODEPTH interaction coefficients suggest that interspecific effects are weaker than intraspecific effects in this system ($\hat{\alpha}<1$, dashed horizontal line, Fig. \ref{Fig:fig3}b), measuring the net biodiversity effect relative to average monoculture yields should significantly inflate any estimate of ecosystem functioning. As would be expected, only when average scaled interaction effects are strong ($\hat{\alpha}>1$), do the simulations suggest that net biodiversity effect, $\Delta\phi_{_\mathbf{T}}$, will become negative by falling below zero (Fig. \ref{Fig:fig3}b).

To gauge the extent to which standard ecosystem functioning measures may be trivially measuring the degree of ecosystem overyielding that is already expected due to coexistence, we can quantify the redundancy in the standard net biodiversity measure $\Delta\phi_{_\mathbf{T}}$.  One possible heuristic approach could involve measuring the ecosystem change along the vector defined by $\mathbf{T} - \mathbf{B}$ in Fig. \ref{Fig:fig2}b, which is simply Loreau-Hector's net biodiversity effect minus the biodiversity effect based on pairwise interactions: $\Delta\phi_{_\mathbf{T}} - \Delta\phi_{_\mathbf{B}}$. This heuristic measure of triviality or redundancy in the net biodiversity effect can be scaled by dividing it by the net biodiversity effect itself to get the scaled or proportional measure of redundancy, $(\Delta\phi_{_\mathbf{T}} - \Delta\phi_{_\mathbf{B}})/ \Delta\phi_{_\mathbf{T}}$, with values above zero indicating redundancy.

For the simulated communities based on the BIODEPTH interaction coefficients, the proportional or scaled redundancy of the Loreau-Hector net biodiversity effect is positive for all the BIODEPTH diversity levels (Fig. \ref{Fig:fig3}c). Clearly, the high redundancy in measured effects ($>> 0$) suggests that most if not all of the overyielding, as measured by net biodiversity effects in the BIODEPTH experiment, are likely to have been already accounted for by coexistence. Overall, these results suggest that by not accounting for the redundant impact of coexistence, current BEF approaches are likely artificially inflating the positive effect of biodiversity on ecosystem functioning.

\subsection*{Measuring and partitioning biodiversity effects under nonlinearity}
\begin{figure}[ht]
  \centering
\includegraphics[scale=0.45]{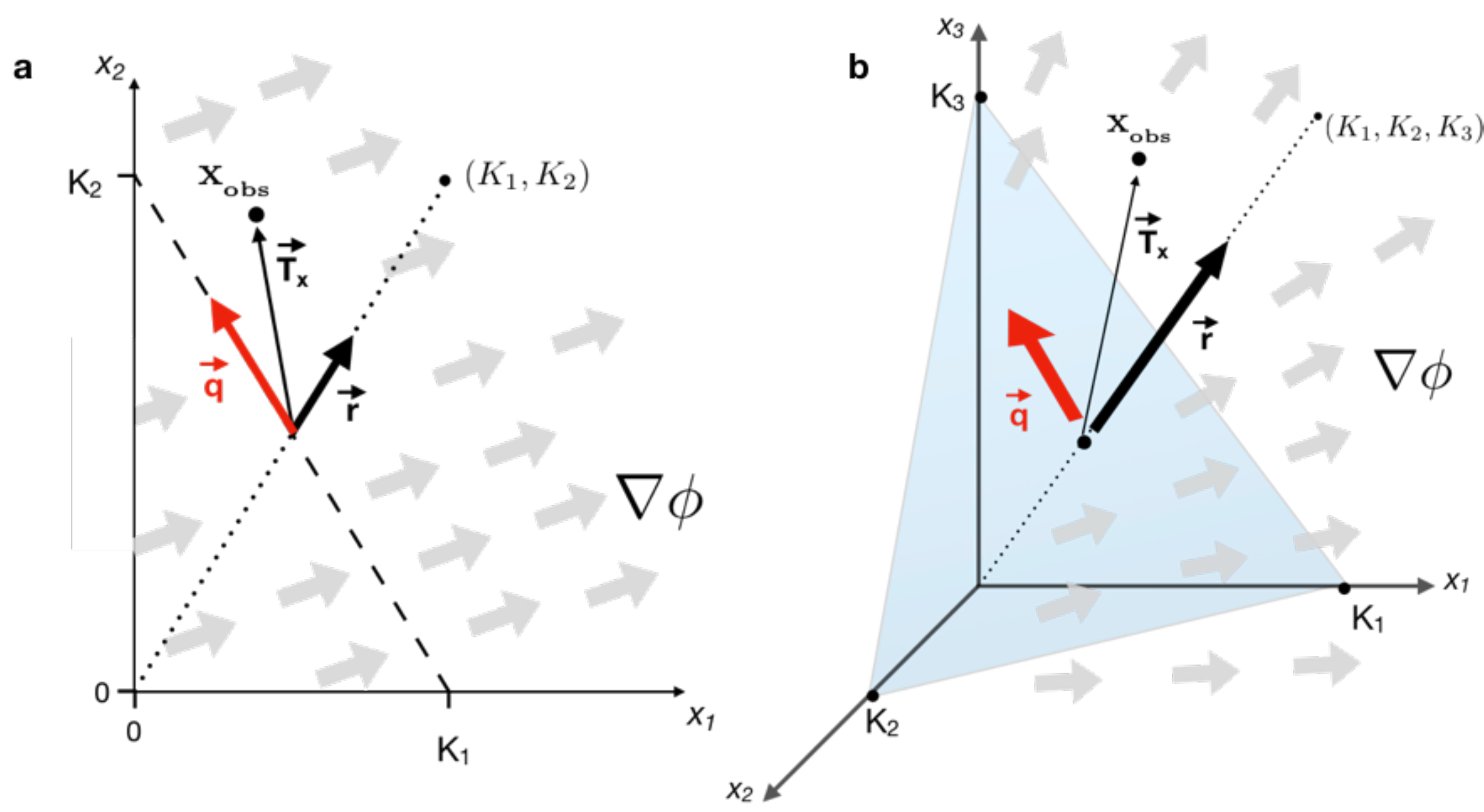}
\caption{Ecosystem changes as changes in underlying community abundance and composition. Visual representation of a (a) two-species and (b) three-species community state space. Gradient of vector field for ecosystem property, $\nabla\phi$, indicated by grey arrows. The displacement vector, $\mathbf{T_x}$, giving the difference in community composition between observed and expected state (center of carrying capacities), can be resolved into two component vectors: one falling on the simplex $\mathbf{q}$ (broad red arrow) giving changes in community composition, and the other along the carrying capacity axis, $\mathbf{r}$ (broad black arrow), gives the change in the average overall community size.
}
\label{Fig:fig4}
\end{figure}

In addition to this quasi-circularity, the underlying theory and methodology of BEF studies are also likely to inflate measured biodiversity effects because they tacitly assume a linear relationship between species abundance and ecosystem functioning in monocultures. To demonstrate the impact of this assumption, we can depict ecosystem properties as functions of the composition and the size of the underlying community. A community's compositional shifts can be followed by tracking its movement through the state space defined by species abundances. The coordinates give the abundances of each species in the community at that point, and the community's state at any given time can be represented by the positional vector $\mathbf{x}=(x_1, x_2, \dotsc,x_n)$. The community space also allows us to define an ecosystem property as a scalar field, where the property is a function of the position with the community state space, $\phi = \phi(\mathbf{x})$. The hyperplane connecting all species carrying capacities now gives the $n$-species community (or  $n$-community) simplex which represents a community surface or space within which shifts in composition represent a zero-sum game (Fig. \ref{Fig:fig4}).

At every point in the community space we can also measure how the ecosystem property is changing by determining the gradient of the ecosystem field at that point, $\nabla\phi$, where $\nabla\phi = \left\langle \frac{\partial \phi}{\partial x_1},\frac{\partial\phi}{\partial x_2}, \dotsc, \frac{\partial\phi}{\partial x_n} \right\rangle $. The ecosystem gradient is thus a conservative vector field indicating the direction and magnitude of the maximum increase in the ecosystem as the community changes in size and composition (Fig. \ref{Fig:fig4}).

If the ecosystem property of each species $i$ in monoculture is a linear function of its abundance, such that $\phi_i=a_ix_i$ (where $a_i$ gives the per capita ecosystem property of $i$), and if we assume (for now) that each species' ecosystem contribution in a mixed community is independent of all other species (no interaction effects), then the aggregate ecosystem property is simply the sum of each species' individual contribution as determined from each species’ ecosystem function in monocultures, $\phi(\mathbf{x}) = \sum_i^n\phi_i(x_i)= a_1x_1+a_2x_2+\dotsb+a_nx_n$. As well, the gradient of the ecosystem field simplifies to a vector field of constant terms $\nabla\phi=\left\langle a_1, a_2,\dotsc,a_n\right\rangle$, indicating that the gradient is  constant and uniform regardless of position.

\subsubsection*{Partitioning biodiversity effects}
Under the typical assumptions of BEF experiments, the expected state of the community based on each species' individual growth rates should be given by each species' carrying capacity divided by $n$:  $\mathbf{x_{_V}}= \frac{1}{n}\mathbf{K}= \frac{1}{n}(K_1, K_2, \dotsc, K_n)$. As with ecosystem measurements, $\frac{1}{n}\mathbf{K}$ provides the reference point from which to measure further transformations in the community that result in the observed community size to depart from expected, $\mathbf{T_x}= \mathbf{x'} - \frac{1}{n}\mathbf{K}$ (where $\mathbf{x'}=(x'_1,\dotsc,x'_n)$ represents the final community state observed).

The $\mathbf{T_x}$ vector, indicating the growth of the community beyond that expected due to the respective fitness of each species (for which carrying capacities serves as a proxy), can itself be resolved into two component vectors that allow us to infer what type of general processes lie behind the observed departures. The component of $\mathbf{T_x}$ that lies along the positional vector connecting the origin to the carrying capacities, $\mathbf{K} = \langle K_1, \dotsc, K_n\rangle$, designated by $\mathbf{r}$, indicates how the community changes due to expansion (or contraction) of the community as a whole, while projecting $\mathbf{T_x}$ onto the simplex that connects all the carrying capacities gives us a vector, $\mathbf{q}$, that describes the changes in the community arising from shifts in species composition possibly due to competition or selection (Fig. \ref{Fig:fig4}). As such, movement along the simplex, $\mathbf{q}$, indicates community compositional shifts that represent a zero-sum game (species replace each other in ratios based on their respective carrying capacities), while $\mathbf{r}$ indicates average fitness growth (or decline) across all species together when competitive or other interaction effects (e.g., facilitation) are symmetrical, allowing $\mathbf{r}$ to be viewed as growth along the niche partitioning axis for the community.

In the simplest case considered here, where the total ecosystem property of a community is the additive sum of each species' ecosystem contribution (no interaction effects), the previously discussed net biodiversity effect, $\Delta\phi_{_\mathbf{T}}$, measured in \textit{ecosystem space} will be equivalent to the total ecosystem change that occurs in the \textit{community state space} along the displacement vector $\mathbf{T_x}$. This is simply the dot product of $\mathbf{T_x}$ and the ecosystem gradient,
\begin{flalign}
  \Delta\phi_{_\mathbf{T}} &= \mathbf{T_x} \cdot \nabla\phi, &{}
\end{flalign}
which gives us
\begin{flalign}\label{eq:partition}
  \Delta\phi_{_\mathbf{T}} &= \mathbf{q}\cdot\nabla\phi + \mathbf{r}\cdot\nabla\phi. &{}
\end{flalign}

The first term in \eqref{eq:partition} gives us the shift in the ecosystem that follows changes in community composition possibly arising from competitive effects between species; the second term gives us the shift in the ecosystem along the niche-partitioning axis $\mathbf{r}$ (note the resemblance of Eq. \eqref{eq:partition} to the expression in the univariate example discussed in the Introduction). Under the assumption of linearity, this ecosystem partitioning is equivalent to Loreau and Hector's partitioning of ecosystem functioning into the ``selection'' and ``complementarity'' effects \citep{loreau.hector_01}, where $\mathbf{q}\cdot\nabla\phi= n\mathrm{Cov}\left[M, \Delta p\right]$, and $\mathbf{r}\cdot\nabla\phi = n\overline{M}\;\overline{\Delta p }$  (and where $\Delta p$ represents the change in the proportion of monoculture for any given species; see Appendix S3).

Although Loreau and Hector suggest that their approach is equivalent to the Price equation, it is better understood as a partitioning of only the transformational component of the Price equation, $\Delta\phi_{_\mathbf{T}}$. Eq. \eqref{eq:partition} yields a clear geometric interpretation of how this ecosystem transformation is occurring as a result of changes in the underlying community, and thus offers a more rigorous and intuitive basis for inferring ecological processes (See S4 for further extensions).

The above vector interpretation of how ecosystem changes relate to shifts in the underlying community not only enables us to easily visualize the partitioning of ecosystem functioning as a result of compositional changes, but also makes self-evident the severe limitations of such partitioning schemes. The formalism of equation \eqref{eq:partition} makes clear that the partitioning of the net biodiversity effect $\Delta\phi_{_\mathbf{T}}$ only holds under strict assumptions of linearity.

If the total ecosystem property in mixtures is due to the additive (non interactive) effect of all species together, then that part of the community shift, $\mathbf{T_x}$,  that is attributable or resolvable to the $\mathbf{q}$ vector will represent the \textit{only} possible source of the ``selection effect'', whereby one is able to speak of species dominating the mixture at the expense of others in a zero-sum game. Similarly, the total community shift that is resolved into the component vector $\mathbf{r}$ will represent the only possible source of the ``complementarity effect'' describing the degree that the total community expands along the niche partitioning axis, such that all species' effects on each other are symmetrical (as with perfect niche complementarity or facilitation). Since, in this special case, these community shifts are the only sources of ecosystem effects, then for the labels ``selection'' and ``complementarity'' to be meaningful within the \textit{ecosystem space}, the compositional shifts in the \textit{community} space have to uniquely (bijectively) map onto the ecosystem shifts apportioned by the Loreau-Hector partitioning, such that the ecosystem effects measured by the LH approach can \textit{only} be attributable to compositional shifts parallel to $\mathbf{q}$ and $\mathbf{r}$, and vice versa.

Under nonlinearity, where monoculture yields are no longer directly proportional to abundance ($\phi_i \neq a_i x_i$), the effects apportioned by the LH method will no longer \textit{uniquely correspond} to the compositional shifts represented by vectors $\mathbf{q}$ and $\mathbf{r}$ in the community space, which, again, are the \textit{only} meaningful sources of selection and complementary (Appendix S3). Only under linear assumptions can the net biodiversity effect, $\Delta\phi_{_\mathbf{T}}$,  be partitioned into effects arising \textit{solely} from shifts in the underlying community, and not from the confounding effects arising from each species' nonlinear ecosystem-abundance relationship in monocultures. Thus, linearity ensures that the Loreau-Hector partitioning will allow us to infer selection and complementary effects purely from measurements within ecosystem space (for proof see Appendix S3.3):
\begin{flalign}\label{eq:partition_equiv}
   n\mathrm{Cov}\left[M, \Delta p\right] &= \mathbf{q}\cdot\nabla\phi \;\;\;\;\;\;\;\textnormal{(Selection\ effect)} &{}\\
   n\overline{M}\;\overline{\Delta p } &= \mathbf{r}\cdot\nabla\phi \;\;\;\;\;\;\;\textnormal{(Complementarity\ effect)} &{}
\end{flalign}

Furthermore, under nonlinearity the community compositional shifts represented by the component vectors $\mathbf{q}$ and $\mathbf{r}$ will not consistently map to any ``selection'' or ``complementarity'' effects in the ecosystem space (Appendix S3.3 for proof). This can be seen by the fact that under nonlinearity, community shifts represented by a component vector, say $\mathbf{q}$, will result in different ecosystem changes depending on the location of $\mathbf{q}$ in the community state space. That is, unless the property gradient is constant due to linearity (such that $\nabla\phi = \langle a_1, a_2,\dotsc, a_n\rangle$), movement along any two (distinct) equal sized parallel vectors, $\mathbf{q_1}$ and $\mathbf{q_2}$ representing identical \textit{community} shifts, will not necessarily result in the same \textit{ecosystem} change, i.e, $ \int_{\mathbf{q_1}} \nabla\phi \cdot d\mathbf{q} \neq \int_{\mathbf{q_2}} \nabla\phi \cdot d\mathbf{q}$.

\subsubsection*{Nonlinearity and the net biodiversity effect}
In addition to the difficulties nonlinearity poses to the development of any partitioning schemes that attempt to infer or ascribe causal effects, it will also likely result in spurious measurements of biodiversity effects \textit{in general}. If the community ecosystem gradient $\nabla\phi(\mathbf{x})$ is not constant due to nonlinear ecosystem-abundance relationships in monocultures (note we are still assuming that the aggregate ecosystem property in mixtures are the additive effects of all species), then any given shift in ecosystem properties (such as the net biodiversity effect along $\mathbf{T}$) will represent not only effects arising from changes in community composition but also from the confounding effects of nonlinearity in each species' individual ecosystem-abundance relationship.

\begin{figure}[h!]
  \centering
\includegraphics[scale=0.48]{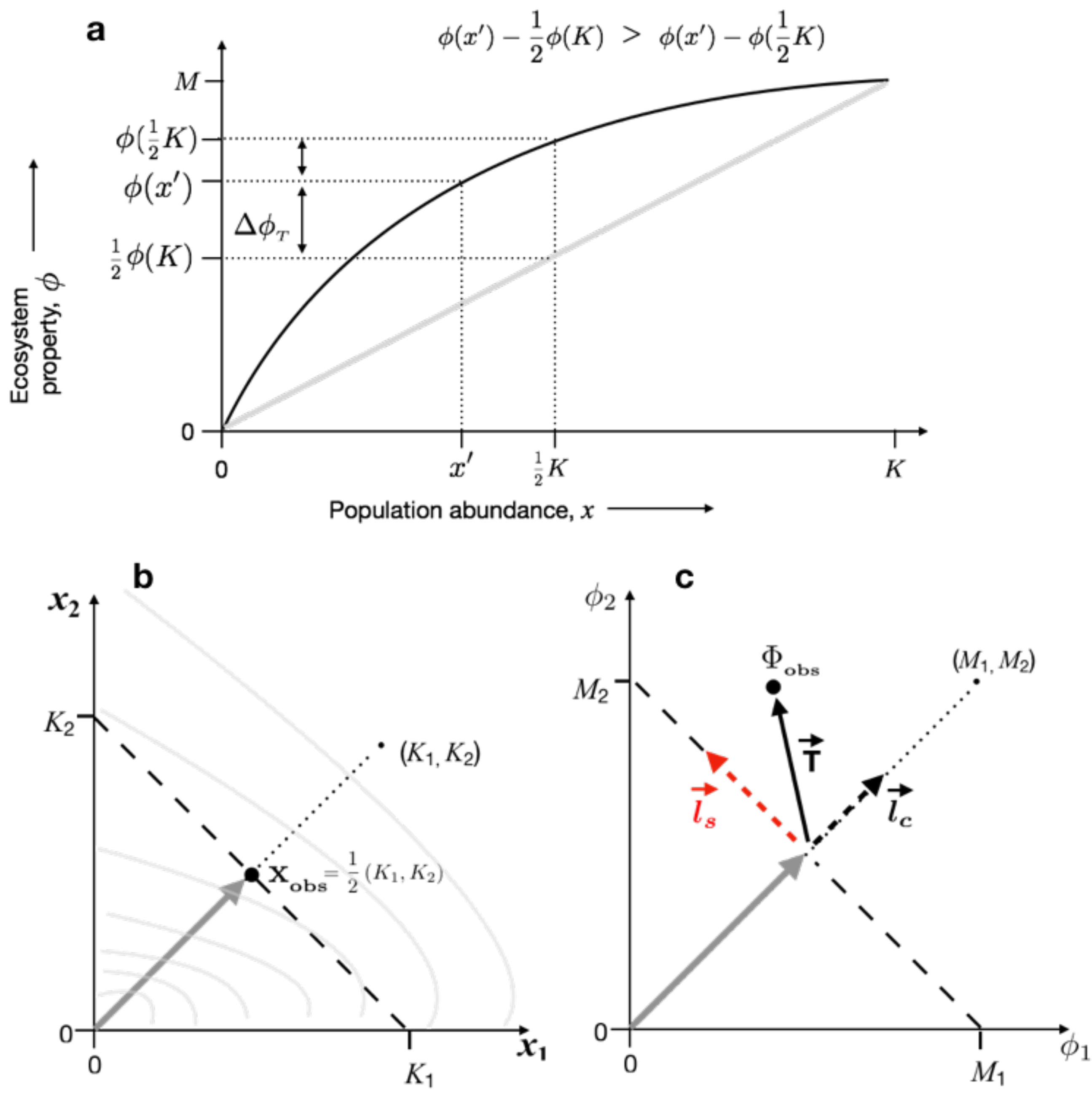}
\caption{Spurious measurements of biodiversity effects. (a) Illustration of how Jensen's inequality leads to inflated biodiversity effects for concave functions. (b) Example in community space of the expected state of a perfectly neutral (randomly assembled) 2-species community at the simplex centroid (no transformational shift $\mathbf{T_x}$).  Grey contour lines represent level surfaces of the ecosystem scalar field (curvature indicates ecosystem property is nonlinear function of community state). (c) The shift  $\mathbf{T}$ observed within the ecosystem space results in a spurious measurement of biodiversity effect ($\Delta \phi_{_\mathbf{T}}>0$), and ``selection'' and ``complementarity'' effects, represented by $\mathbf{T}$'s component vectors, $\mathbf{l_s}$ and $\mathbf{l_c}$, respectively.}
\label{Fig:fig5}
\end{figure}

Nonlinearity in ecosystem-abundance relationships is almost certainly the rule, not the exception. Even simple properties like plant biomass have been known to display strikingly nonlinear responses to changes in population density, as has been extensively documented in the self-thinning literature \citep[see for example \textit{Yoda's power law};][]{yoda.kira.ea_63,westoby_84,enquist.brown.ea_98}. What's more, given that many (perhaps most) ecosystem properties of interest in nature are likely to exhibit concave -- or at least saturating -- response curves with respect to species abundance \citep{scrosati_06,demirezen.aksoy.ea_07,stachova.fibich.ea_13,li.weiner.ea_16}, the biodiversity effect sizes measured in current BEF studies are likely to be artificially and significantly inflated. This is a simple consequence of Jensen's inequality and is likely to inflate the measured net biodiversity effect $\Delta\phi_{_\mathbf{T}}$ by ensuring that the average monoculture yields that serve as the baseline for measuring net biodiversity effects will be consistently lower than the actual ecosystem property that corresponds to the community state where all species equally share the niche (i.e., $\frac{1}{n}\langle K_1, K_2, \dotsc, K_n\rangle$; see Fig. \ref{Fig:fig5}a).

The potential for concave functions to artificially inflate biodiversity measurements can be conceptually illustrated with a neutral two-species community where both species are at the midpoint of their carrying capacities (equivalent to a species distribution expected from a random binomial sampling; Fig \ref{Fig:fig5}b). For a given species with a concave response curve, the ecosystem change associated with the difference in abundance between that expected, $\frac{1}{2}K$, and the observed, $x'$, will be $\phi(x')- \phi(\frac{1}{2}K)$. However the ecosystem shift measured from the midpoint of the monoculture yield to the observed, $\Delta\phi_{_\mathbf{T}} = \phi(x') - \frac{1}{2}\phi(K)$ (Fig \ref{Fig:fig5}a), which serves as the basis for measuring net biodiversity effect, will consistently be larger than the actual ecosystem change measured relative to the point where both species equally share the niche, $\phi(\frac{1}{2}K)$,
\begin{flalign}
  \Delta\phi_{_\mathbf{T}} &> \phi(x')- \phi(\frac{1}{2}K). &{}
\end{flalign}

If either species individually has a concave ecosystem functional response, then not only will there be a positive biodiversity effect measured in this community (vector $\mathbf{T}$, Fig. \ref{Fig:fig5}c), but also potentially selection and complementarity effects (along the vectors $\mathbf{l_s}$ and $\mathbf{l_c}$ shown in Fig. \ref{Fig:fig5}c, respectively). Since both species are at abundances expected in a neutral community (or from random sampling), and since the aggregate ecosystem properties observed are those expected from their single species response curves (i.e., no species interaction effects), it is meaningless here to talk of there being a `biodiversity effect' on ecosystem functioning, let alone a `complementarity' or `selection' effect. Any such effects measured would be artefacts of the nonlinear functional responses of individual species instead of biodiversity.

\begin{figure}[h!]
  \centering
    \includegraphics[width=0.85\textwidth]{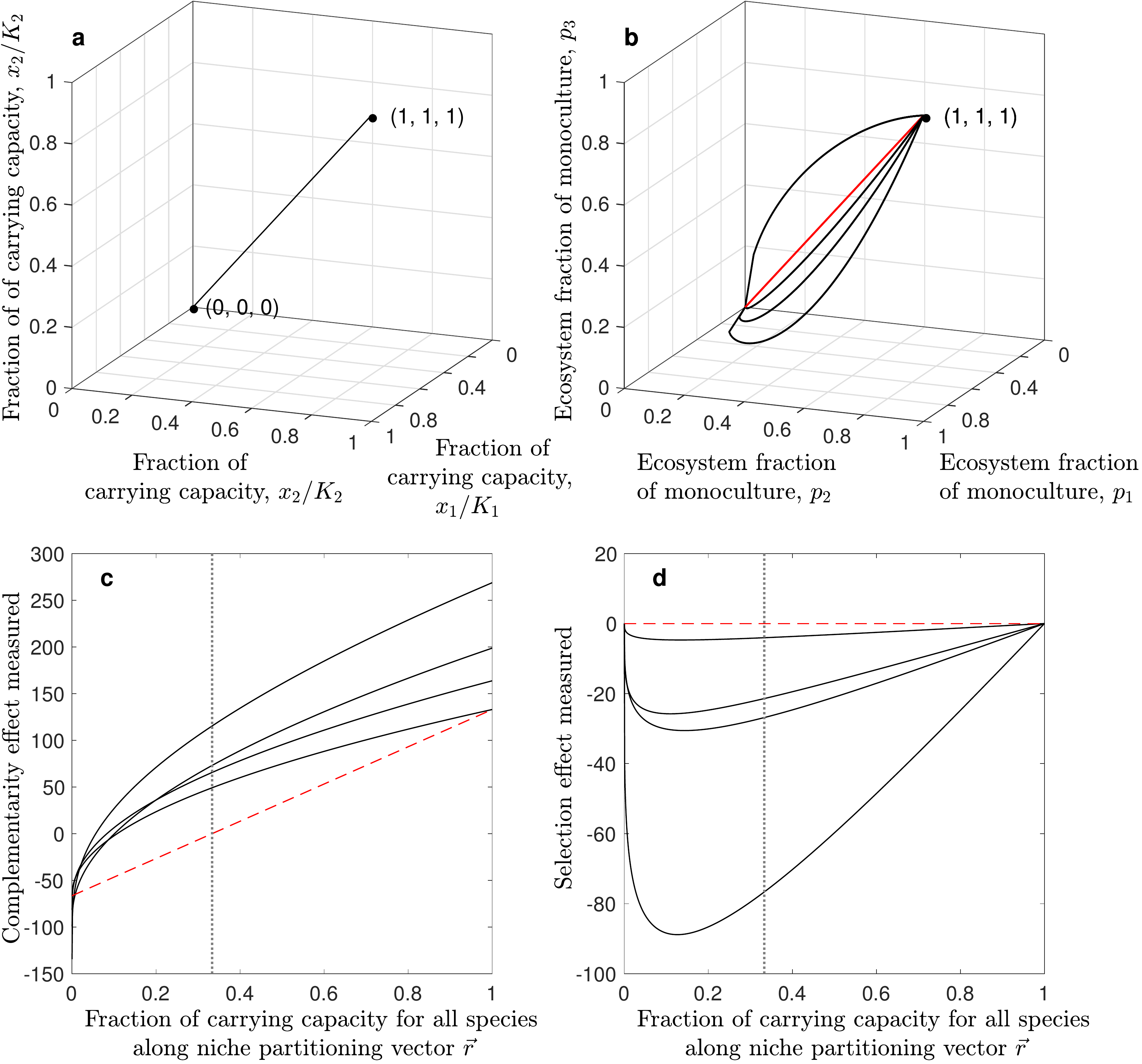}
    \caption{Nonlinearity and the partitioning of ecosystem functioning. Example of a 3-species system. (a) Complementarity axis connecting the origin to the carrying capacities of all species in the community space (abundances given as proportion of carrying capacity, $x_i/K_i$). If ecosystem property is a linear function, then movement along this line results in only a complementarity effect being measured, but no selection effect. (b) Observed ecosystem trajectory (as proportion of monoculture yield) in the ecosystem space for four communities with non-linear ecosystem-abundance curves as one moves up the complementarity axis in panel (a). Dashed red line indicates trajectory when function is linear. (c) Complementarity effects and (d) selection effects measured for  curves shown in (b). In (c) and (d) dashed red lines show expected complementarity or selection effect when ecosystem functions are linear; dotted vertical line marks the 1/3 fraction of $K$  where the community state is at the simplex centroid (i.e., point where both ecosystem effects are expected to vanish). All nonlinear communities show significant departures from the expected complementarity and selection effect, including when the community is at the centroid of the community simplex where effect sizes should be zero (intersection of dashed and vertical line).}
    \label{Fig:fig6}
\end{figure}

We can further demonstrate the spurious measurements of biodiversity effects by tracking the aggregate ecosystem properties of various three-species communities along the niche partitioning axis (Fig. \ref{Fig:fig6}a), where the aggregate ecosystem property is the additive (non-interactive) total of individual species effects (see Appendix S2 for specific parameters). If individual species response curves are nonlinear, the aggregate ecosystem property at the community level will not follow the niche partitioning (or complementarity) axis in the ecosystem space (red dashed line, Fig. \ref{Fig:fig6}b). Because communities are constrained to exist along the niche partitioning axis within the community space, and because the community ecosystem property does not involve interaction effects, we would expect no selection effect and any biodiversity effect detected should be due solely to the role of complementarity, which itself should be positive only when all species are above the abundance fraction of 1/3. Yet, for all four communities shown, we detect striking departures from expectations, with each community showing a notable selection effect and all communities displaying a positive complementarity effect even when species are significantly below the threshold abundance fraction of 1/3 (Fig. \ref{Fig:fig6}c,d). Hence, spurious biodiversity, selection and complementarity effects could emerge solely due to the nonlinearity in individual species’ ecosystem functioning-abundance relationships in monocultures.

Although we demonstrated the necessary conditions allowing ecosystem partitioning by excluding the effects of species interactions on community properties in mixtures, our vector partitioning in Eq. \eqref{eq:partition} nevertheless holds when the aggregate community-level properties involve interactive effects (Appendix S5). Effects arising from species compositional changes and species interactions (including nonlinear interaction effects) are community-level effects that can be reasonably attributed to diversity. However, BEF studies essentially confound the effects species composition and species interactions have on aggregate ecosystem properties with the effects of nonlinear ecosystem-abundance relationships in individual species monocultures.

Put another way, much of the biodiversity effects observed in high diversity communities may reflect the nonlinear ecosystem responses within individual species monocultures, and not the effects of species diversity itself. These nonlinear effects are unaccounted for when monoculture yields are used to determine the expected or reference yield $\phi_{_V}$. More so, these same nonlinear effects within individual species render the Loreau and Hector partitioning incapable of separating out the contributions of niche partitioning and selection/competition towards the overall biodiversity effect. We suspect that disentangling these nonlinear effects may be impossible using current BEF experimental design approaches.

\section*{Conclusions}

Our critique has primarily centered on the logic and mathematics of scientific inference in BEF research: both the problematic logic of the BEF approach to measuring ecosystem functioning (due to its built-in circularity), and the flawed mathematics of the LH partitioning scheme, which continues to play a foundational role in the analysis of BEF data. As we demonstrated above, this latter mathematical flaw due to nonlinearity not only undermines the inferential power of the Loreau-Hector statistical partitioning scheme, but also renders the net biodiversity effect itself as a spurious measurement.

It should be clear here that we are not challenging the generally positive biodiversity-ecosystem functioning relationship recorded by researchers for a quarter of a century. The question is not whether the measurement of this positive relationship (whether by using raw or net biodiversity measures) is real or not, it is how meaningful such measurements and patterns obtained are. Given that coexistence theory already leads us to expect that overyielding of ecosystem properties would be a natural outcome of coexistence (all things being equal), it is hardly surprising then that BEF researchers should have discerned an overall positive biodiversity effect using metrics and experimental designs formulated with a naive null hypothesis that does not account for built-in effects incumbent upon coexistence itself.

Similarly, it is useful to note that the arithmetic underlying the Loreau and Hector partitioning method, although meaningless, is still formally correct. It is possible to use their partitioning approach purely as a descriptive metric for shifts in the ecosystem space without anchoring them in changes in the underlying community. However, the effects measured by their partitioning approach would then merely become a phenomenological description of ecosystem change devoid of any ecological insight; an arbitrary (and inferentially pointless) scoring and partitioning of the surface phenomena of observed ecosystem changes.  Moreover, the use of the terms ``complementarity'' and ``selection'' to describe the ecosystem quantities apportioned by the Loreau-Hector method become arbitrary labels stripped of their original biological meaning.

Without knowing the functional relationship between ecosystem properties and the underlying community, conducting experiments that are restricted to measuring ecosystem properties alone will not allow us to infer underlying ecological processes through any partitioning of the ecosystem changes as done by Loreau and Hector, or establish the effect biodiversity may have on ecosystem properties using current experimental approaches. Again, what is under question is not the ``reality'' of the phenomenological descriptions or measurements made in nature (or their arbitrary partitioning into smaller units), but rather, whether such measurements can truly allow us to infer the explanatory causes that have been claimed by theorists and experimentalists for decades.

These serious theoretical and methodological flaws in current experimental approaches underscore the importance of conducting BEF experiments that do not infer species fitness from monoculture yields or try to infer ecological processes solely from patterns of movement in ecosystem phase space. Our use of pairwise interactions to measure BEF effects builds on the precedence of using pairwise interactions to infer community assembly mechanisms \citep{kraft.godoy.ea_15}, and thus not only offers an opportunity to begin developing more rigorous approaches to studying ecosystems, but also of anchoring such studies in nearly a century of basic ecological theory.

\section*{Acknowledgments}This research was supported by a grant from the National Science Foundation (OCE-1458158) to TCG.

\paragraph{Statement of authorship:} PP designed the study, developed the theory, analytical framework, approach/methods and critique; PP ran the simulations, analysed the results, and wrote the paper; TG helped to analyse the results and write the paper.

\bibliography{pillai_gouhier_2018ms_BEF_biblio}

\bibliographystyle{ecology_letters2}



\section*{Supplementary material}

# S1 Partitioning ecosystem change and the Price equation

The positional vector $\mathbf{\Phi_V}$ represents the expected state (or coordinates) of the system after a single time step due to the effects of differential growth or reproduction of the component species independent of each other. The expected total value of the ecosystem property at this state, $\phi_\mathbf{V}$, arising due to variation in growth or fitness of each component species is simply $\phi_\mathbf{V} = \sum_i^n w_i \phi_i = n\mathbf{E}[w\phi]$, where $\phi_i$ represents the initial ecosystem property contribution of the $i$th species, and $w_i$ its corresponding fitness.

The difference in the ecosystem property between this expected value, $\phi_\mathbf{V}$, and the observed value, $\phi_\mathbf{obs}$, represents an additional ecosystem shift that arises from transformational evolution or changes in the system ($\Delta\phi_\mathbf{T}$) that are not simply reducible to variation in species growth (such as changes due to ecologial interactions, frequency dependent selection, enviromental effects, etc.). If we consider the final or observed ecosystem contribution of each $i$ species to be $\phi'_i$, such that $\phi_\mathbf{obs} = \sum_i^n \phi'_i$ for an $n$-species community, then:

$$
\begin{aligned}
\Delta\phi_\mathbf{T} &= \phi_\mathbf{obs} - \phi_\mathbf{V} \\
&= \sum_{i=1}^n \phi'_i - \sum_i^n w_i \phi_i \\
&= \sum_{i=1}^n w_i \left(\hat{\phi}'_i - \phi_i\right) \\
&= n\mathbf{E}\left[w\Delta\hat{\phi}\right],
\end{aligned}
\qquad (S1)
$$

where $\hat{\phi}'_i = \phi'_i/w_i$. Thus the final observed measure of an ecosystem property after a bout of change is simply the sum of both the effects that variation in the growth rates of component species have on the ecosystem, and an additional ecosystem shift due to the effects of further transformational changes, such as those due to ecological interactions, that are not associated with variation in fitness or growth of individual species:

$$\phi_\mathbf{obs} = \phi_\mathbf{V} + \Delta\phi_\mathbf{T}. \qquad (S2)$$

This is simply a modified form of the Price equation. In biodiversity-ecosystem studies, expectations based on the monoculture yields serve as a multiplicative factor (or a proxy for fitness) that gives the variational growth rate of each system part or component (i.e., species), and the ecosystem property due to variation in growth, $\phi_\mathbf{V}$, is assumed to simply be the average of the monoculture yields.

Alternatively, we can subtract the initial value of the ecosystem property, $\phi_\mathbf{init} = \sum_i^n \phi_i$, from both the final observed value and the the expected value due to variational selection, on both sides of Eq. (S2), to get the total net change in the system, $\Delta\phi_\mathbf{total}$, after a single

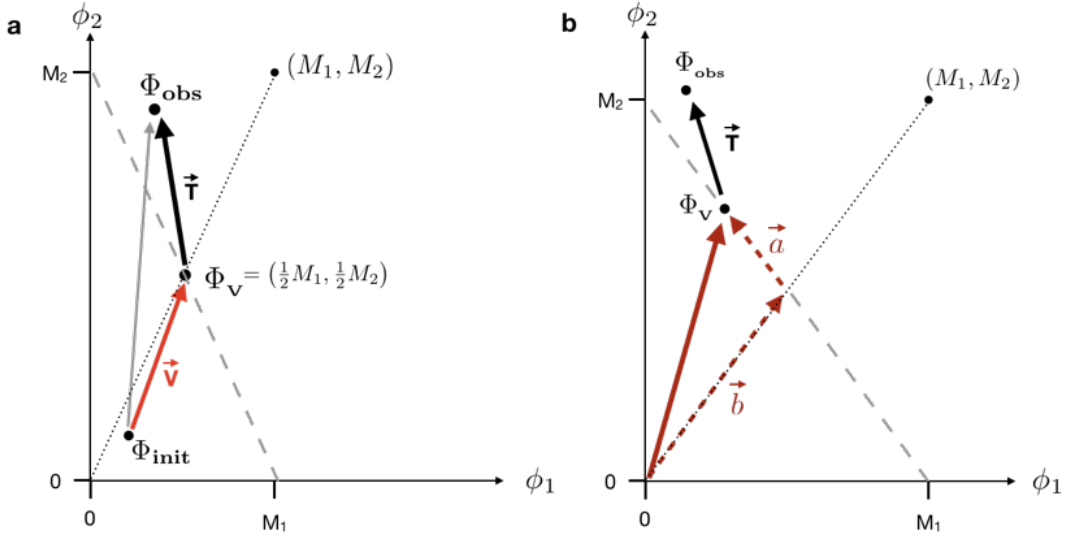

**Figure S1:** Visual representation of ecosystem change. (a) Variational and transformational changes in an ecosystem as vectors in BEF experiments (Eq. (S3)). (b) If the expected state of the ecosystem due to variation in growth, $\Phi_V$, is represented as a positional vector (red solid arrow), then the Price equation is a version of Eq. (S2) that only includes the component of this vector along the simplex from the centroid: $\Delta\phi_\mathbf{a} + \Delta\phi_\mathbf{T}$, where $\Delta\phi_\mathbf{a} = n\mathrm{Cov}[w,\phi]$.

time step:

$$\phi_{\mathbf{obs}} - \phi_{\mathbf{init}} = (\phi_\mathbf{V} - \phi_{\mathbf{init}}) + \Delta\phi_\mathbf{T}$$

$$\Delta\phi_{\mathbf{total}} = \left[\sum_{i=1}^n w_i\phi_i - \sum_{i=1}^n \phi_i\right] + \Delta\phi_\mathbf{T}$$

$$= \left[n\mathbf{E}[w\phi] - n\bar{\phi}\right] + \Delta\phi_\mathbf{T}$$

$$= \left[n\mathrm{Cov}[w,\phi] + n(\bar{w}-1)\bar{\phi}\right] + \Delta\phi_\mathbf{T}$$

$$\Delta\phi_{\mathbf{total}} = \Delta\phi_\mathbf{V} + \Delta\phi_\mathbf{T}. \tag{S3}$$

We can see here how the total ecosystem change can be partitioned into the effects arising from changes due to variational evolution and transformational evolution of the system. If the displacement vectors **V** and **T** represent the variational and transformational components of the total change in state that the system undergoes in a time step, then both $\Delta\phi_\mathbf{V}$ and $\Delta\phi_\mathbf{T}$ measure the actual changes in ecosystem properties along the **V** and **T** vectors, respectively.

The Price equation itself is usually expressed in a form obtained by simply subtracting the expression $n\bar{w}\bar{\phi}$ (which gives the change along the vector **b** in Fig. S1b) from both $\phi_{\mathbf{obs}}$

and $\phi_{\mathbf{V}}$ on each side of Eq. (S2),

$$\begin{aligned}
\left(\phi_{\mathbf{obs}} - n\bar{w}\bar{\phi}\right) &= \left(\phi_{\mathbf{V}} - n\bar{w}\bar{\phi}\right) + \Delta\phi_{\mathbf{T}} \\
n\left(\bar{\phi}' - \bar{w}\bar{\phi}\right) &= n\mathrm{Cov}\left[w, \phi\right] + \Delta\phi_{\mathbf{T}} \\
\bar{w}\left(\frac{\bar{\phi}'}{\bar{w}} - \bar{\phi}\right) &= \mathrm{Cov}\left[w, \phi\right] + \mathbf{E}\left[w\Delta\hat{\phi}\right] \\
\bar{w}\Delta\bar{\hat{\phi}} &= \mathrm{Cov}\left[w, \phi\right] + \mathbf{E}\left[w\Delta\hat{\phi}\right].
\end{aligned} \tag{S4}$$

## S2 Supplementary Methods

### S2.1 Simulations using BIODEPTH data

We used all the available BIODEPTH data (Spehn *et al.*, 2005) that included both the total and relative biomasses for each species in a given diversity treatment (diversity levels: 2, 3, 4, 8, 11, 12, 14). We then calculated the average ecosystem functioning (aggregate biomass) using the standard measure $\Delta\phi_{\mathbf{T}}$ over three years.

All available two-species mixtures were used to calculate interaction coefficients, $\alpha_{i,j}$. In order for us to measure ecosystem functioning using our metric based on pairwise interactions, $\Delta\phi_{\mathbf{B}}$, we would need to have available all the possible pairwise interactions for all species in the given treatment (i.e., $\binom{n}{2}$ pairwise interactions for an $n$-species treatment). However, the species combinations available in the BIODEPTH two-species mixture treatments were insufficient to directly test ecosystem functioning (aggregate biomass) relative to our baseline.

We therefore simulated in MATLAB an artificial species pool (1000 species), each with a monoculture yield randomly drawn from a log normal distribution, using the mean ($\mu = 5.2701$) and standard deviation ($\sigma = 0.9297$) calculated from the available monoculture yields for all species in the treatments listed above. We then calculated the interaction coefficients, $\alpha_{i,j}$, for all the pairs in the 2-species treatment. In order to calculate the average and variance of the pairwise interactions we scaled the coefficients in the following manner:

$$\widehat{\alpha}_{i,j} = \alpha_{i,j} \times \frac{M_j}{M_i}. \tag{S5}$$

Scaling the coefficients enabled us to control for the effects that variation in monoculture yields have on the coefficients, and thus allowed the interaction effects between different pairs to be directly comparable. For example, cases where $\hat{\alpha}_{i,j} = 1$ would now consistently indicate the equivalence of intraspecific and interspecific competition (i.e., a neutral interaction) regardless of the actual monoculture yields. The mean and standard deviation of $\hat{\alpha}_{i,j}$ from BIODEPTH was found to be $\mu = 0.3528$ and $\sigma = 0.6822$, respectively.

At each diversity level studied ($n = 3, 4, 8, 11, 12$, and $14$) we sampled 1000 randomly assembled communities from the species pool, then averaged the total biodiversity effect (calculated using our metric $\Delta\phi_{\mathbf{B}}$) over the total number of samples for each diversity level. The results shown in Fig. 3a are the average of 100 such simulations.

The experiment was both a proof of concept for our approach, and a demonstration (using existing data) of how this system of guild species is characterized by relatively low average interaction/competition coefficients, indicating a strong tendency for the effects of intraspecific competition to outweigh those of interspecific competition (as would be predicted from co-existence theory).

## S2.2 Nonlinear functions used in Figure 6

In Fig. 6 we plotted four 3-species communities, where each species had a nonlinear ecosystem-relationship in monoculture. The functional form of the ecosystem-abundance relationship for each species $i$ was a power function, $\phi_i = a_i x_i^{t_i}$, and the aggregate ecosystem property for the community was simply the sum of power functions, $\phi(\mathbf{x}) = \sum_i a_i x_i^{t_i}$. Each species in all four communities had identical coefficients $a_i$, as well as the same carrying capacities, $K_i$: $a_i = 5$ and $K_i = 150$ for all $i$.

The exponent $t_i$ varied for the four communities as follows: community 1, $t_1 = 0.7$, $t_2 = 0.5$, $t_3 = 0.25$; community 2, $t_1 = 0.3$, $t_2 = 0.5$, $t_3 = 0.75$; community 3, $t_1 = 0.18$, $t_2 = 0.4$, $t_3 = 0.85$; community 4, $t_1 = 0.4$, $t_2 = 0.5$, $t_3 = 0.6$.

# S3 Geometric interpretation of ecosystem partitioning

The Loreau and Hector (2001) partitioning of ecosystem functioning purportedly measures both what they call the "complementarity effect" and "selection effect" of ecosystem change as a function of biodiversity. Expressions for both these effects can be obtained by a trivial partitioning of the transformational term of the Price equation, $\Delta \phi_\mathbf{T}$ (Eq. (S1)). For the $i$th species' monoculture yield $M_i$ and its proportional change in ecosystem functioning $\Delta p_i$, both effects are defined as follows:

Selection effect:

$$n \times \text{Cov}\,(M, \Delta p) = \sum_{i=1}^{n} \Delta p_i M_i - \overline{\Delta p} \sum_{i=1}^{n} M_i, \tag{S6}$$

Complementarity effect:

$$n \times \overline{M}\,\overline{\Delta p} = \overline{\Delta p} \sum_{i=1}^{n} M_i. \tag{S7}$$

These effects can also be visualised geometrically in an $n$-species ecosystem state space (Fig. S2). In an ecosystem space where the ecosystem properties are defined as proportions of the monoculture yields, $(p_i = \frac{\phi_i}{M_i})$, the simplex is the plane connecting all the axes at 1, while the centroid of the simplex is simply $(\frac{1}{n}, \frac{1}{n}, \ldots, \frac{1}{n})$. If the observed state of the ecosystem is $(p'_1, p'_2, \ldots, p'_n)$, then the displacement vector $\mathbf{\Delta p}$ describing the change in ecosystem functioning is $\langle \Delta p_1, \Delta p_2, \ldots, \Delta p_n \rangle = (p'_1, p'_2, \ldots, p'_n) - (\frac{1}{n}, \frac{1}{n}, \ldots, \frac{1}{n})$.

The vector projection of $\mathbf{\Delta p}$ onto the simplex, which we designate here as $\mathbf{\Delta p_q}$, gives the proportional change (relative to the monoculture yield) in the system due to the "selection effect" defined above. Let $\mathbf{z}$ be the unit vector normal to the simplex, where

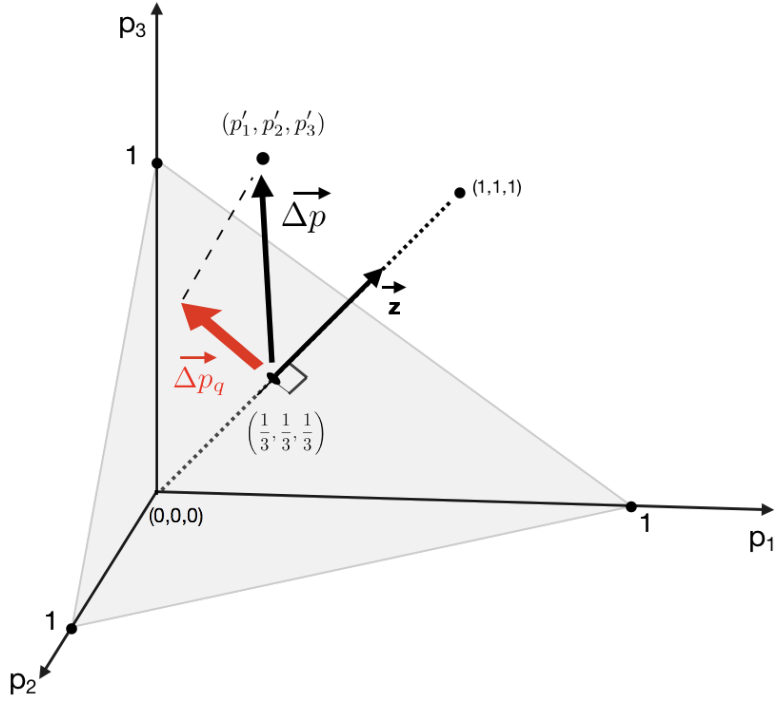

**Figure S2:** Vector interpretation of ecosystem partitioning in a 3-species community.

$\mathbf{z} = \left\langle \frac{1}{\sqrt{n}}, \frac{1}{\sqrt{n}}, \ldots, \frac{1}{\sqrt{n}} \right\rangle$, then

$$\boldsymbol{\Delta p_q} = \boldsymbol{\Delta p} - (\boldsymbol{\Delta p} \cdot \mathbf{z})\,\mathbf{z}$$

$$= \left\langle \Delta p_1, \Delta p_2, \ldots, \Delta p_n \right\rangle - \left( \frac{1}{\sqrt{n}} \sum_{i=1}^{n} \Delta p_i \right) \left\langle \frac{1}{\sqrt{n}}, \frac{1}{\sqrt{n}}, \ldots, \frac{1}{\sqrt{n}} \right\rangle$$

$$= \left\langle \Delta p_1 - \overline{\Delta p},\ \Delta p_2 - \overline{\Delta p}, \ldots,\ \Delta p_n - \overline{\Delta p} \right\rangle \tag{S8}$$

The dot product of $\boldsymbol{\Delta p_q}$ and $\mathbf{M}$, where $\mathbf{M} = \langle M_1, M_2, \ldots, M_n \rangle$ then gives the total selection effect in (S6):

$$\boldsymbol{\Delta p_q} \cdot \mathbf{M} = \sum_{i=1}^{n} \Delta p_i M_i - \overline{\Delta p} \sum_{i=1}^{n} M_i. \tag{S9}$$

Similarly, $\boldsymbol{\Delta p_r}$ designates the vector giving the proportional ecosystem change along $\mathbf{z}$ due to the "complementarity effect". Since $\boldsymbol{\Delta p} = \boldsymbol{\Delta p_q} + \boldsymbol{\Delta p_r}$, we have $\boldsymbol{\Delta p_r} = \left\langle \overline{\Delta p}, \ldots, \overline{\Delta p} \right\rangle$. Then, as in Eq. (S6), the total ecosystem change due to the "complementarity effect" is

$$\boldsymbol{\Delta p_r} \cdot \mathbf{M} = \overline{\Delta p} \sum_{i=1}^{n} M_i. \tag{S10}$$

## S3.1 Ecosystem properties as linear functions of community composition

If the ecosystem property is a linear function of the position within the community space, such that $\phi = \phi(\mathbf{x}) = a_1 x_1 + a_2 x_2 + \cdots + a_n x_n$, then it follows that in general $\Delta \phi_i = \Delta x_i \frac{\partial \phi}{\partial x_i}$, and more specifically that $M_i = K_i \frac{\partial \phi}{\partial x_i}$ and $\Delta p_i = \frac{\Delta \phi_i}{M_i} = \frac{\Delta x_i}{K_i}$. This leads to the following:

$$\begin{aligned}
\mathbf{\Delta p_q} \cdot \mathbf{M} &= \sum_{i=1}^{n} \Delta p_{q_i} M_i \\
&= \sum_{i=1}^{n} \Delta p_{q_i} \left( K_i \frac{\partial \phi}{\partial x_i} \right) = \sum_{i=1}^{n} (\Delta p_{q_i} K_i) \frac{\partial \phi}{\partial x_i} \\
&= \sum_{i=1}^{n} q_i \frac{\partial \phi}{\partial x_i}.
\end{aligned} \quad (S11)$$

When the initial point of the $\mathbf{\Delta p}$ vector is the centroid of the simplex, $(\frac{1}{n}, \ldots, \frac{1}{n})$, then the $i$th element of $\mathbf{q}$ is

$$q_i = x'_i - \overline{p'} K_i, \quad (S12)$$

where $x'_i$ is the observed abundance of the $i$th species. Combining (S9) and (S11) shows how, under conditions of linearity, movement along the simplex in the community state space describes the community compositional shifts underlying Loreau and Hector's "selection effect",

$$n \times \text{Cov}(M, \Delta p) = \mathbf{q} \cdot \nabla \phi. \quad (S13)$$

We can similarly show how the complementarity effect arises from the ecosystem shifts attributable to the the $\mathbf{r}$ vector component of the community's movement in state space,

$$\begin{aligned}
\mathbf{\Delta p_r} \cdot \mathbf{M} &= \sum_{i=1}^{n} \Delta p_{r_i} M_i = \sum_{i=1}^{n} (\Delta p_{r_i} K_i) \frac{\partial \phi}{\partial x_i} \\
&= \sum_{i=1}^{n} r_i \frac{\partial \phi}{\partial x_i}.
\end{aligned} \quad (S14)$$

Here the $i$th element of $\mathbf{r}$ is given by $r_i = \overline{\Delta p}\, K_i$. Combining (S10) and (S14) then gives the so-called "complementarity effect",

$$n \times \overline{M}\, \overline{\Delta p} = \mathbf{r} \cdot \nabla \phi. \quad (S15)$$

## S3.2 Proof confirming that q lies in the plane paralllel to $n$-community simplex

Let $\mathbf{z}$ be a vector normal to the simplex (hyper)plane in the community state space. Vector $\mathbf{q}$ will be in a plane parallel to the simplex if and only if the following holds:

$$\mathbf{q} \cdot \mathbf{z} = 0. \quad (S16)$$

7

The $n$-species community (or $n$-community) simplex connecting all carrying capacities, $K_i$, is an $n$–1-dimensional hyperplane that satisfies the following $n$ equations:

$$\left\langle \left(1-\frac{1}{n}\right)K_1,\ -\frac{1}{n}K_2,\ \ldots,\ -\frac{1}{n}K_i,\ \ldots,\ -\frac{1}{n}K_n \right\rangle \cdot \mathbf{z} = 0$$

$$\left\langle -\frac{1}{n}K_1,\ \left(1-\frac{1}{n}\right)K_2,\ \ldots,\ -\frac{1}{n}K_i,\ \ldots,\ -\frac{1}{n}K_n \right\rangle \cdot \mathbf{z} = 0$$

$$\vdots$$

$$\left\langle -\frac{1}{n}K_1,\ -\frac{1}{n}K_2,\ \ldots,\ \left(1-\frac{1}{n}\right)K_i,\ \ldots,\ -\frac{1}{n}K_n \right\rangle \cdot \mathbf{z} = 0$$

$$\vdots$$

$$\left\langle -\frac{1}{n}K_1,\ -\frac{1}{n}K_2,\ \ldots,\ -\frac{1}{n}K_i,\ \ldots,\ \left(1-\frac{1}{n}\right)K_n \right\rangle \cdot \mathbf{z} = 0 \tag{S17}$$

Since the solution to the above equations is simply

$$z_i = \left(\overline{zK}\right) \times \frac{1}{K_i}, \quad \text{for all } 1 \le i \le n, \tag{S18}$$

then $\mathbf{z}$ can be any vector parallel to $\langle \frac{1}{K_1}, \frac{1}{K_2}, \ldots, \frac{1}{K_n}\rangle$. Or in normalized (unit vector) form,

$$\mathbf{z} = \frac{1}{\sqrt{\sum_i^n \frac{1}{K_i^2}}} \left\langle \frac{1}{K_1}, \frac{1}{K_2}, \ldots, \frac{1}{K_n} \right\rangle. \tag{S19}$$

Combining Eqs. (S12) and (S18) with condition (S16),

$$\mathbf{q} \cdot \mathbf{z} = 0$$

$$\sum_{i=1}^n x_i z_i - \overline{p} \sum_{i=1}^n K_i z_i = 0$$

$$\sum_{i=1}^n x_i \frac{1}{K_i} = \overline{p} \sum_{i=1}^n K_i \frac{1}{K_i}$$

$$n\overline{p} = \overline{p}n. \tag{S20}$$

Therefore, since Eq. (S16) holds, $\mathbf{q}$ is in the $n$-community simplex (or a parallel hyperplane).

## S3.3 Demonstration of how LH partitioning holds only under linearity

The Loreau-Hector (or any similar) partitioning scheme can only infer the operation of community-level processes like selection and complementarity if the function that maps changes in the community space to the corresponding shifts observed in the ecosystem space

is linear and bijective.

**Claim 1**: If the total community-level change can be partitioned into the sum of individual processes (such as selection and complementarity), then the corresponding total ecosystem change in mixtures (including the LH net biodiversity effect) will be the sum of the (mutually exclusive) ecosystem effects that are solely attributable to distinct underlying community-level processes, if and only if the function $\psi$, that maps composition and abundance changes in community space to the corresponding shifts in ecosystem state space, is linear.

**Claim 2**: This linear mapping must be bijective if one wishes to measure and attribute the effects obtained from a partitioning of observed ecosystem changes to various underlying community-level processes. Specifically, an effect measured in an ecosystem partitioning scheme can only be attributed to, or used to infer a given community-level change if the ecosystem effect measured uniquely arises from this specific community-level shift.

**Claim 3**: Since BEF partitioning experiments implicitly measure ecosystem functioning relative to the null assumption that each species ecosystem contribution is independent of all others (no interactive effects), then so long as the above linear condition holds, the condition of bijectivity is automatically satisfied.

Below we demonstrate the proof of these claims.

**Definitions**

**(1)** Recall that for a vector-valued function or map $\mathcal{L}$ to be linear, the following must hold for any two arbitrary vectors **a** and **b** in a vector space $A$, and for scalar constant $c$:

$$\mathcal{L}(\mathbf{a} + \mathbf{b}) = \mathcal{L}(\mathbf{a}) + \mathcal{L}(\mathbf{b}), \tag{S21}$$
$$\mathcal{L}(c\,\mathbf{a}) = c\,\mathcal{L}(\mathbf{a}). \tag{S22}$$

**(2)** If **x** and **y** represent displacement vectors in the community and ecosystem space, respectively, then let $\psi$ be the vector-valued function mapping shifts in the community space to shifts in ecosystem space, $\psi(\mathbf{x}) = \mathbf{y}$.

In an $n$-species system, any change in community composition and abundance represented by the displacement vector **x** will result in a change in the ecosystem state of the system represented by displacement vector **y**. The vector-valued function $\psi$ will map shifts in the community space to shifts in ecosystem space, such that $\psi(\mathbf{x}) = \mathbf{y}$, where $\psi : \mathbb{R}^n \to \mathbb{R}^n$.

**Claim 1: Condition of linearity**

Let us assume that the displacement vector in the community space **x** can be represented as the sum of two vectors **q** and **r**, such that $\mathbf{x} = \mathbf{q} + \mathbf{r}$, where $\mathbf{q} \in Q$ and $\mathbf{r} \in R$ for subsets $Q \subset \mathbb{R}^n$ and $R \subset \mathbb{R}^n$.

We will also assume that the subsets $Q$ and $R$ represent sets of vectors associated with distinct ecological processes in the community space (say selection and complementarity), and are thus, except for intersecting at the **0** vector, non-overlapping subsets of $\mathbb{R}^n$.

Now, in order for us to split the total ecosystem shift given by any $\mathbf{y} = \psi(\mathbf{x})$ exclusively into the ecosystem effects arising from $\mathbf{q}$ and $\mathbf{r}$ displacements in the community space, it is required that

$$\mathbf{y} = \psi(\mathbf{q} + \mathbf{r}) = \psi(\mathbf{q}) + \psi(\mathbf{r}). \tag{S23}$$

If Eq. (S23) holds, the total ecosystem shift $\mathbf{y}$ can be considered the sum of mutually exclusive ecosystem effects $\psi(\mathbf{q})$ and $\psi(\mathbf{r})$, arising from community shifts represented by $\mathbf{q}$ and $\mathbf{r}$, respectively. However, from condition (S21) we know that Eq. (S23) will only hold if $\psi$ is a linear mapping $\mathbb{R}^n \to \mathbb{R}^n$.

This means that an ecosystem change will only appear as the additive total of effects arising from distinct community changes when the mapping from the community to the ecosystem space is linear. The corollary of this is that if the map is not linear, the community-level shifts or processes will not appear as distinct effects in the ecosystem space, but may be confounded together; the result of which is that no simple partitioning scheme at the ecosystem surface level will be capable of separating out the confounded effects.

**Claim 2: Condition of bijectivity**

We have established that only if $\psi$ is a linear function can an ecosystem change $\mathbf{y}$ be considered as the sum of $\psi(\mathbf{q})$ and $\psi(\mathbf{r})$, for example, the sum of effects arising exclusively from selection and complementarity. However, knowing that an observed ecosystem change can be attributed to different, mutually exclusive underlying effects does not mean a given ecosystem partitioning scheme will allow us to measure or attribute observed ecosystem changes to such effects. We will now show that if, in addition to being linear, $\psi$ is a bijective function, then it will be possible to have a partitioning scheme that will allow us to associate partitioned effects measured in ecosystem space *uniquely* to compositional shifts in community space.

Imagine, that in order to measure the ecosystem effects of $\mathbf{q}$ and $\mathbf{r}$, a partitioning scheme was devised that allows us to resolve the ecosystem displacement vector $\mathbf{y}$ as the sum of two vectors $\mathbf{u}$ and $\mathbf{v}$, such that $\mathbf{y} = \mathbf{u} + \mathbf{v}$. Under the condition of linearity we can, by using Eq. (S23), then claim

$$\mathbf{u} = \psi(\mathbf{q}), \tag{S24}$$
$$\mathbf{v} = \psi(\mathbf{r}). \tag{S25}$$

From Eqs. (S24)-(S25) it would appear that our partitioning of $\mathbf{y}$ into $\mathbf{u}$ and $\mathbf{v}$ would allow us to attribute the partitioned ecosystem shifts exclusively to the effects of $\mathbf{q}$ and $\mathbf{r}$. However, if the function $\psi$ is not injective such that it allows two distinct types of community shifts, say $\mathbf{q_1} \in Q$ and $\mathbf{r_1} \in R$, to map to the same vector in the ecosystem space, say $\mathbf{u_1}$, then our linear mapping will not necessarily allow the ecosystem effects obtained from partitioning to be exclusively attributed to given community-level effects.

Similarly, it might be possible that the function $\psi$ is not surjective, such that not every $\mathbf{y}$ in the ecosystem space has been mapped to from a displacement $\mathbf{x}$ in the community space. This means that measured ecosystem effects may not be attributable to any underlying community shifts whatsoever.

If however the linear map $\psi$ is bijective, then $\psi$ is both an injective and surjective map, such that for all ecosystem changes **y** measured there is a *unique* (one and only one) **x** in the community space. This means that if Eqs. (S24)-(S25) hold, then partitioned ecosystem effects **u** and **v** can be used to unambiguously attribute ecosystem effects to a given community-level process.

**Claim 3: Under linearity the bijectivity condition is automatically satisfied**

In BEF experiments, if the map $\psi$ from community to ecosystem space is linear, then $\psi$ is also bijective (or invertible). Since expectations in BEF experiments are based on species ecosystem responses in monocultures (where species' ecosystem properties are not affected by interaction effects from other species), the null assumption is that the ecosystem property or contribution of each species in mixtures will be unaffected by other species' ecosystem contributions. That is, the null expectation for the $i^{th}$ species' ecosystem contribution will be $\phi_i = f(x_i)$. When $f(x_i)$ is linear, then $\phi_i = a_i x_i$, for constant $a_i$.

This means that if $\psi$ is a linear transformation in the form of a matrix $D$, then $D$ is a diagonal matrix with $a_i$ elements along the main diagonal, and zeros elsewhere. Since $D$ is a diagonal matrix it is also invertible, such that the inverse $D^{-1}$ exists. Thus, for any linear transformation $D: q \mapsto u$ relating a community-level processes to a partitioned ecosystem effect, the inverse mapping also exists, $D^{-1}: u \mapsto q$, uniquely relating an ecosystem effect with an underlying community process.

All of this implies that once we know that the ecosystem responses of all species in monocultures are linear, then we will also know that the ecosystem effects of community-level processes can be accurately partitioned and measured at the ecosystem level because the bijective condition is automatically satisfied.

# S4 Supplementary results: Extensions of vector partitioning approach when ecosystem property is a linear function of abundance

If the vector $\mathbf{B_x}$ represents the difference within the community state space between the community observed and the one predicted from pairwise interactions, then under linear assumptions, the biodiversity effect that we defined earlier, $\Delta\phi_{\mathbf{B}}$, can be expressed as $\Delta\phi_{\mathbf{B}} = \mathbf{B_x} \cdot \nabla\phi$. This means that we can partition the biodiversity effect $\Delta\phi_{\mathbf{B}}$ in a similar manner to Eq. (5) in the main article by simply resolving $\mathbf{B_x}$ into its corresponding **q** and **r** vector components (Fig. S3a).

Furthermore, if we project the ecosystem gradient $\nabla\phi$ onto the simplex, the angle, $\theta$, between the resulting vector field **s** (where $\mathbf{s} = \nabla\phi - (\nabla\phi \cdot \mathbf{z}) \cdot \mathbf{z}$ for unit vector **z** normal to simplex) and **q** will give a measure of correlation ($\cos\theta$) describing the degree to which the species composition of the community is being driven in the direction of increasing ecosystem functioning (i.e., is there a tendency for competition or selection to favour species making greater contributions to a given ecosystem property (Fig. S3b)).

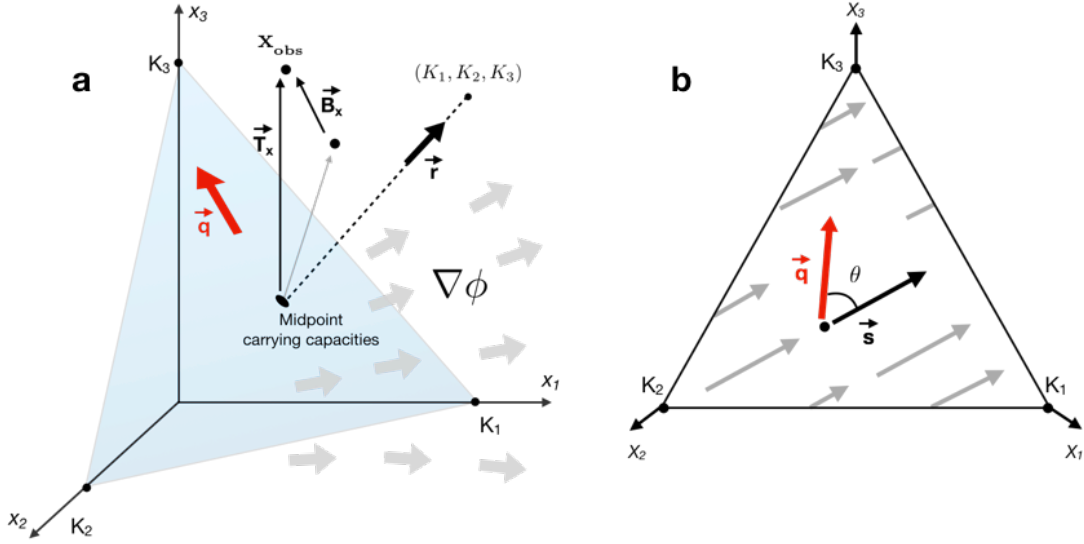

**Figure S3:** Measuring biodiversity effects through changes in underlying community abundance and composition. (a) Displacement vector $\mathbf{B_x}$, showing the community state's shift from the expected based on pairwise ineteractions. Resolving $\mathbf{B_x}$ into its component $\mathbf{r}$ and $\mathbf{q}$ vectors shows the departure in both community composition and total size from that expected based on pairwise interactions. (b) Projecting the ecosystem gradient onto the simplex gives the direction and magnitude of maximum ecosystem change, $\mathbf{s}$, along the community simplex (example shown here for a 3-species case); cosine of the angle $\theta$ between $\mathbf{q}$ and $\mathbf{s}$ provides the correlation measuring how observed shifts in community composition correspond to an increase in ecosystem property.

## S5 Aggregate ecosystem properties arising from both changing abundances and interaction effects in mixtures

The total difference between the aggregate ecosystem property observed in a system ($\phi_{\mathbf{obs}} = \sum_i \phi_i'$) and that expected ($\phi_{\mathbf{exp}} = \sum_i \phi_i$) is

$$\Delta \phi = \phi_{\mathbf{obs}} - \phi_{\mathbf{exp}} = \sum_{i=1}^{n} \phi_i' - \sum_{i=1}^{n} \phi_i$$

$$= \sum_{i=1}^{n} \left( \frac{\phi_i'}{\phi_i} - 1 \right) \phi_i. \tag{S26}$$

Note that if $\Delta \phi = \Delta \phi_{\mathbf{T}}$, then we can consider $\phi_{\mathbf{exp}}$ as being equivalent to $\phi_{\mathbf{V}}$ in Eq (S2).

If $x_i$ and $x_i'$ are the abundances at the expected and observed states, respectively, then let $a_i = \phi_i/x_i$ and $a_i' = \phi_i'/x_i'$, where $a_i$ is the per capita ecosystem contribution of species $i$

12

at the expected state, $\phi_i$, as predicted from monocultures, and $a'_i$ is the per capita ecosystem contribution of species $i$ that is actually observed in mixtures. In the main paper we assumed that there was no interaction effects on per capita ecosystem properties ($\phi_i = \phi_i(x_i)$), and that, under linearity, changes in the aggregate ecosystem properties of mixtures arises solely due to changing species abundances, such that $a'_i = a_i$. Now we relax this assumption and allow the per capita properties to change in mixtures due to interaction effects,

$$\Delta\phi_{\mathbf{T}} = \sum_{i=1}^{n} \left( \frac{x'_i}{x_i} \frac{a'_i}{a_i} - 1 \right) x_i a_i$$
$$= \sum_{i=1}^{n} \left( \lambda_i \, \mu_i - 1 \right) x_i a_i. \tag{S27}$$

In the above expression $\lambda_i$ is the growth factor representing the scale of the $i^{th}$ species' abundance shift in mixture relative to the expected, $x'_i = \lambda_i x_i$. If all species exhibit linear ecosystem response curves in monocultures then the factor $\mu_i$ gives the scale by which the per capita ecosystem property of $i$ is observed to have been magnified in the mixture. If there are no interaction effects on per capita properties in mixtures then $\mu_i = 1$. We will consider the expression $\hat{x}' = \mu_i x'_i = \mu_i(\lambda_i x_i)$ as giving the observed ecosystem impact of the species $i$ in mixtures relative to the monoculture.

In general we can consider $\nabla\phi = \left\langle \frac{\partial \phi_1(x_1)}{\partial x_1}, \ldots, \frac{\partial \phi_n(x_n)}{\partial x_n} \right\rangle$ as the ecosystem gradient for an $n$-species community based on monoculture functional responses, where $\phi_i(x_i)$ is the monoculture ecosystem property of species $i$. If the ecosystem functional response in monocultures for each species is linear ($\phi_i = a_i x_i$ for constant $a_i$), then the ecosystem gradient of the community will be represented by the the vector field $\nabla\phi = \langle a_1, a_2, \ldots, a_n \rangle$. We can then rewrite Eq (S27) as

$$\Delta\phi_{\mathbf{T}} = \sum_{i=1}^{n} \left[ \left( \mu_i \, \lambda_i - 1 \right) x_i \right] a_i$$
$$= \sum_{i=1}^{n} \hat{T}_{x_i} a_i$$
$$= \hat{\mathbf{T}}_{\mathbf{x}} \cdot \nabla\phi. \tag{S28}$$

The displacement vector $\hat{\mathbf{T}}_{\mathbf{x}}$ shows how each species' impact on aggregate ecosystem properties changes in mixture relative to that expected from monoculture yields. Now we are no longer just considering the impact of species abundance alone on ecosystem properties, but the total impact that each species has on the aggregate ecosystem property due to both abundance changes along wth the scaling of the per capita ecosystem property represented by $\mu_i$. The total change in species' ecosystem impact when in mixtures, $\hat{\mathbf{T}}_{\mathbf{x}}$ (Eq (S28)), can be partitioned in a similar manner to Eq (5) in the main paper,

$$\Delta\phi_{\mathbf{T}} = \hat{\mathbf{q}} \cdot \nabla\phi + \hat{\mathbf{r}} \cdot \nabla\phi. \tag{S29}$$

Since the monoculture ecosystem response functions are linear, the changes represented by $\hat{\mathbf{T}}_{\mathbf{x}}$ are solely due to community-level effects. This is because the parameter $\mu$ in the

expression $(\mu_i\,\lambda_i - 1)\,x_i$ only reflects the scaling of per capita ecosystem properties arising from community-level interactions.

Partitioning $\hat{\mathbf{T}}_\mathbf{x}$ tells us the degree to which changes in each species ecosystem impact complement each other or come at each other's expense in a zero-sum game. In the limiting case where $\mu_i = 1$ for all $i$, changes in aggregate properties ascribed to niche partitioning and competition/selection effects are due to changes in species abundances alone, as outlined in the main paper. Alternatively, in the opposite limiting case where species abundances in mixture are at the expected level ($\lambda_i = 1$), effects of niche partitioning and competition arise from the degree each species' per capita property contribution is magnified in mixture.

Take for instance a two-species plant community where both species are at the midpoint of their respective carrying capacities. If both species have a symmetrically positive effect on each other's rate of absorption of some nutrient, then a complementarity effect will be measured where the aggregate rate or stock of nutrient absorbed will increase along the niche partitioning axis. If, on the other hand, one species monopolizes or absorbs the nutrient perfectly at the expense of the other species, then the per capita rate of absorption of each species will be scaled in such a way as to appear as movement along the simplex line connecting the two carrying capacities in the community space (defined by each species' ecosystem impact), which will then be observed as a perfectly constrained shift along the monoculture simplex in the ecosystem space.

Now let us consider Eq. (S27) when all monoculture ecosystem responses are non-linear (since per capita property values are no longer constant in monocultures, we will use $a_i^*$ to distinguish the per capita property at the expected state from other states):

$$\Delta\phi_{\mathbf{T}} = \sum_{i=1}^{n} \big[\,(\mu_i\,\lambda_i - 1)\,x_i\big]a_i^*$$
$$= \Big\langle (\mu_1\,x_1' - x_1),(\mu_2\,x_2' - x_2),\ldots,(\mu_n\,x_n' - x_n)\Big\rangle \cdot \Big\langle a_1^*, a_2^*,\ldots, a_n^*\Big\rangle. \tag{S30}$$

Equation (S30) is similar to Eq. (S28), except now we can no longer factor out the ecosystem gradient based on monocultures since $\langle a_1^*, a_2^*, \ldots, a_n^*\rangle \neq \nabla\phi$. More importantly, the first vector on the right hand side of Eq. (S30) no longer represents changes solely ascribable to community-level processes; specifically, the parameter $\mu$ now confounds both the effects on ecosystem properties of community-level interactions in mixtures with that of the non-linear responses associated with single species monocultures.

We can still partition the vector of changes represented by the first vector on the R.H.S of Eq. (S30), but because everything is now confounded together such a partitioning no longer provides an ecologically meaningful measure, as was the case with Loreau and Hector's approach. In fact, if instead of factoring out $x_i\,a_i^*$ in the above expression we factor out the total monoculture yield $M_i$, then partitioning Eq. (S30) will produce the original Loreau and Hector formula.

14



\end{document}